\documentclass[10pt]{article}
\usepackage{sectsty}
\usepackage{extsizes}
\usepackage[super,sort&compress,comma]{natbib} 
\usepackage[version=3]{mhchem}
\usepackage[left=2cm, right=2cm, top=1.785cm, bottom=2.0cm]{geometry}
\usepackage{balance}
\usepackage{mathptmx}
\usepackage{graphicx} 
\usepackage{lastpage}
\usepackage[format=plain,labelfont=bf,labelsep=space]{caption}
\usepackage{float}
\usepackage{fancyhdr}
\usepackage{fnpos}
\usepackage[english]{babel}
\addto{\captionsenglish}{%
  
}
\usepackage[usenames,dvipsnames]{xcolor}
\usepackage{setspace}
\usepackage{hyperref}

\usepackage{epstopdf}
\usepackage{physics}

\definecolor{cream}{RGB}{222,217,201}

\begin{document}
\setstretch{1.25}
\thispagestyle{plain}

\makeatother

\renewcommand{\thefootnote}{\fnsymbol{footnote}}
\renewcommand\footnoterule{\vspace*{1pt}%
\color{cream}\hrule width 3.5in height 0.4pt \color{black}\vspace*{5pt}}

\title{\textbf{Variability in X-ray induced effects in \ce{[Rh(COD)Cl]2} with changing experimental parameters}} 
\date{\vspace{-8ex}}
\maketitle

\noindent\large{Nathalie K. Fernando,\textit{$^{a}$}$^{\ast}$ Hanna L. B. Bostr{\"o}m,\textit{$^{b}$} Claire A. Murray,\textit{$^{c}$} Robin L. Owen,\textit{$^{c}$} Amber L. Thompson,\textit{$^{d}$} Joshua L. Dickerson,\textit{$^{e}$} Elspeth F. Garman,\textit{$^{f}$} Andrew B. Cairns,\textit{$^{g}$} and Anna Regoutz\textit{$^{a}$}$^{\ast}$} \\

\noindent X-ray characterisation methods have undoubtedly enabled cutting-edge advances in all aspects of materials research. Despite the enormous breadth of information that can be extracted from these techniques, the challenge of radiation-induced sample change and damage remains prevalent. This is largely due to the emergence of modern, high-intensity X-ray source technologies and growing potential to carry out more complex, longer duration \textit{in-situ} or \textit{in-operando} studies. The tunability of synchrotron beamlines enables the routine application of photon energy-dependent experiments. This work explores the structural stability of \ce{[Rh(COD)Cl]2}, a widely used catalyst and precursor in the chemical industry, across a range of beamline parameters that target X-ray energies of 8~keV, 15~keV, 18~keV and 25~keV, on a powder X-ray diffraction synchrotron beamline at room temperature. Structural changes are discussed with respect to absorbed X-ray dose at each experimental setting associated with the respective photon energy. In addition, the X-ray radiation hardness of the catalyst is discussed, by utilising the diffraction data at the different energies to determine a dose limit, which is often considered in protein crystallography and typically overlooked in small molecule crystallography. This work not only gives fundamental insight into how damage manifests in this organometallic catalyst, but will encourage careful consideration of experimental X-ray parameters before conducting diffraction on similar radiation-sensitive organometallic materials. \\


\begin{NoHyper}
\footnotetext{\textit{$^{a}$~Department of Chemistry, University College London, 20 Gordon Street, London, WC1H~0AJ, UK.}}
\footnotetext{\textit{$^{b}$~Max Planck Institute for Solid State Research, Heisenbergstra{\ss}e 1, 70569 Stuttgart, Germany.}}
\footnotetext{\textit{$^{c}$~Diamond Light Source Ltd, Diamond House, Harwell Science and Innovation Campus, Didcot, Oxfordshire, OX11~0DE, UK.}}
\footnotetext{\textit{$^{d}$~Chemical Crystallography, Chemistry Research Laboratory, University of Oxford, South Parks Road, Oxford OX1~3QR, UK.}}
\footnotetext{\textit{$^{e}$~MRC Laboratory of Molecular Biology, Francis Crick Avenue, Cambridge Biomedical Campus, Cambridge CB2 0QH, UK.}}
\footnotetext{\textit{$^{f}$~Department of Biochemistry, Dorothy Crowfoot Hodgkin Building, University of Oxford, South Parks Road, Oxford, OX1~3QU, UK.}}
\footnotetext{\textit{$^{g}$~Department of Materials, Imperial College London, Royal School of Mines, Exhibition Road, SW7~2AZ, UK.}}
\footnotetext{\textit{$^{\ast}$~Corresponding authors: a.regoutz@ucl.ac.uk, nathalie.fernando.13@ucl.ac.uk.}}

\end{NoHyper}



\section{Introduction}
X-ray radiation, essential to a vast array of modern characterisation methods, can induce sample change as a result of inelastic photon--matter scattering interactions. The study of radiation damage has seen significant advancements over the past two decades, particularly focused on macromolecular crystallography (MX). Notable advancements in MX-based radiation damage include the categorisation of structural damage, the reproducible order and progression of damage to specific chemical bonds typical for extended biological molecules, the calculation of absorbed radiation dose, and the development of innovative damage mitigation strategies. Detailed insight into radiation damage in protein crystallography is discussed in a large number of publications, including several comprehensive review articles.~\cite{Holton2009, Garman2010, Taberman2018, Garman2017}\par 
Despite the leaps in biological radiation damage research, systematic studies of the impact of X-ray radiation and its progression in small molecule crystals are rare. This has become a pressing issue in recent years, due to the exponential rise in X-ray source brilliance, both for synchrotron and laboratory sources.~\cite{Morgan2018, Christensen2019, Kubin2018, George2008} A deeper understanding of how certain properties of the incoming photon beam, such as the energy, contribute to sample damage in small molecule systems at room temperature, is required to better plan X-ray experiments with regards to sample stability under irradiation. This information is crucial since experimental setups of small molecule systems in powder diffraction studies typically differ from protein diffraction, where the protein crystal/solution is irradiated and measured in the presence of a solvent or water.\par

The use of higher X-ray energies in characterisation experiments is known to reduce the photoelectric absorption and thereby sample damage.~\cite{Arndt1984, Nave2005} However, the elastic scattering cross section responsible for diffraction also decreases with increasing incident X-ray energy. Several investigations into the effect of photon energies on radiation damage in protein crystallography exist both for 100 K and room temperature experiments, with the aim of determining an optimum photon energy for diffraction experiments. Early reports focused on finding an ideal energy at which damage is minimised and encouraged the use of shorter X-ray wavelengths, typically less than 0.9~\AA, corresponding to photon energies greater than 13.8~keV, to reduce radiation damage in proteins.~\cite{Arndt1984, Polikarpov1997} This was later corroborated by M\"{u}ller~\textit{et al.}~who reported that damage in an organic light atom system was reduced in the energy (wavelength) range of 12.4~keV~(1.0~\AA) to 20.7~keV~(0.6~\AA) at 100~K.~\cite{ Muller2002} In contrast, later reports reported no evidence of energy dependence in diffraction intensity across certain energy ranges. For instance, Weiss~\textit{et al.}~saw no notable difference from 6.4~keV~(2.0~\AA) to 12.4~keV~(1.0~\AA) at 100~K.~\cite{Weiss2005} Shimizu~\textit{et al}.~extended the previously studied energy range, by irradiating lysozyme crystals at 100~K from 6.1~keV to 33.0~keV.~\cite{Shimizu2007} The authors also note that no clear dependence of reciprocal space damage, notably mosaicity and B-factors, on photon energy could be found when using the absorbed dose as the metric. The work by Shimizu~\textit{et al.}~highlighted the often overlooked importance of considering not only the radiation-effects of varying energies, but also the absorbed dose.\par
Dose, defined as the X-ray energy absorbed per unit mass, is used to quantify levels of sample X-ray exposure. It is a convolution of experimental factors such as photon flux, energy, absorption, and photoionisation, with sample properties (elements, unit cell \textit{etc}.). One limitation of the existing literature is the focus on the energy dependence of the global damage observed, rather than on the significance of specific local damage. Initial studies by Homer~\textit{et al.}~indicate an energy dependence of site-specific damage to sulfur environments in lysozyme crystals cooled to 100~K, with the electron density around the cysteine sulfur atoms decreasing faster at 14~keV than at 9~keV.~\cite{Homer2011}\par
At higher incident energies, the much lower quantum efficiency of beamline Si detectors, in addition to the notable drop in photon flux, contributes to an overall, indirect energy dependence of radiation damage. The issue of decreased detector efficiency at higher energies has long been a cause for concern.~\cite{Donath2013} In recent years, the development of CdTe detectors with higher quantum efficiency at incident energies greater than 15~keV for macromolecular crystallography experiments, has introduced the possibility of using higher energies than those typically used at the synchrotron, to mitigate radiation damage.~\cite{Dickerson2019, Storm2021} Dickerson~\textit{et al.}~determined a theoretical optimum photon energy of 26~keV which offers the greatest diffracted signal per unit dose, when considering the Quantum Efficiency (QE) of a CdTe detector. \par


A figure of merit often used to describe the stability of a macromolecular sample under specific measurement conditions is the dose limit. The Henderson limit, $D_{0.5}$, of 20~MGy, is the threshold typically used in the field of MX to assess the dose at which the diffraction intensity of a protein crystal, cooled to 100~K, is reduced by half due to the X-ray induced perturbation of long-range crystal order.~\cite{Henderson1990} The limit was derived by analogy from observations of electron diffraction decay with electron flux density. Subsequent X-ray studies by Owen~\textit{et al.}~have redefined this value to 30~MGy, using a $D_{ln2} \approx D_{0.7}$ upper limit since biological information is found to be compromised at $D_{0.5}$.~\cite{Owen2006} However, this value has been challenged as too high. Atakisi~\textit{et al.}~proposed a local resolution-dependent half-dose as a more robust metric in protein crystallography. The dose limit at $T$ = 100~K was found to increase from 2–3~MGy at 1~\AA~to 30~MGy at 4~\AA. For lysozyme crystals diffracting to a maximum resolution of 1.4~\AA, the half-dose was determined to be 10~MGy.~\cite{Atakisi2019} Teng~\textit{et al.}~also determined a dose threshold of 10~MGy in lysozyme crystals at 100~K.~\cite{Teng2000}\par 
It is worth noting that the definition of dose limits is variable across different X-ray experiment setups and sample environments. For instance, tolerable dose limits in soft X-ray absorption spectroscopy (XAS) experiments of metal compound solid samples tend to be orders of magnitude greater than for a sample in solution.~\cite{Kubin2018} This is attributed to the decrease in electron and radical mobility and the associated drop in recombination effects, which overpowers the diffusion-driven damage in solution samples.~\cite{Kubin2018} Additionally, cryocooling a crystal is well known to extend the dose limit by several orders of magnitude and hence, crystal lifetime.~\cite{Nave2005_2, Liebschner2015} How standard such sample environments are is highly variable, particularly in the case of powder X-ray diffraction, where measurements are more often conducted at room temperature. Furthermore, in non-MX diffraction, samples are generally not in the parent liquor of crystallisation, a factor which can drastically alter the tolerable dose limit of a sample. As there is typically no water in or surrounding samples studied with powder diffraction (except in the case of porous materials), radiolysis of water does not usually occur. As such, any mention of dose limits in previous single crystal X-ray diffraction (SCXRD) studies is not directly comparable with small molecule crystals in powder X-ray diffraction (PXRD). At present, despite the importance for experiment planning, there are very few reported dose limits for small molecule systems in the solid phase.\par 

In this study, a model organometallic catalyst \ce{[Rh(COD)Cl]2}, where COD=1,5-cyclooctadiene, is studied. This system has wide-reaching applications in the chemical industry, for instance in hydroamination and hydrogenation reactions, and also serves as a model framework for similar organometallic structures, related by metal centre, halide or organic ligand.~\cite{Ashfeld2007} A previous radiation damage study of this catalyst by the authors using X-ray diffraction (XRD), X-ray photoelectron spectroscopy (XPS), and density functional theory (DFT), highlighted the differences in sensitivity of damage across the two experimental methods.~\cite{Fernando2021} The present work expands on this study, focusing on PXRD and exploring the influence of the experimental setup on the radiation-induced structural changes occurring in the catalyst. In small molecule experiments, to avoid unfavourable energy ranges (where sample absorption is high) the positions of absorption edges are often considered. The excitation energy is then tuned accordingly. Typically, experimental factors beyond photon energy, such as X-ray dose and dose rate are rarely considered, despite evidence from MX suggesting a strong influence of these factors on the quality of diffraction data acquired.\par

\begin{figure}[ht]
\centering
\includegraphics[width=0.485\textwidth]{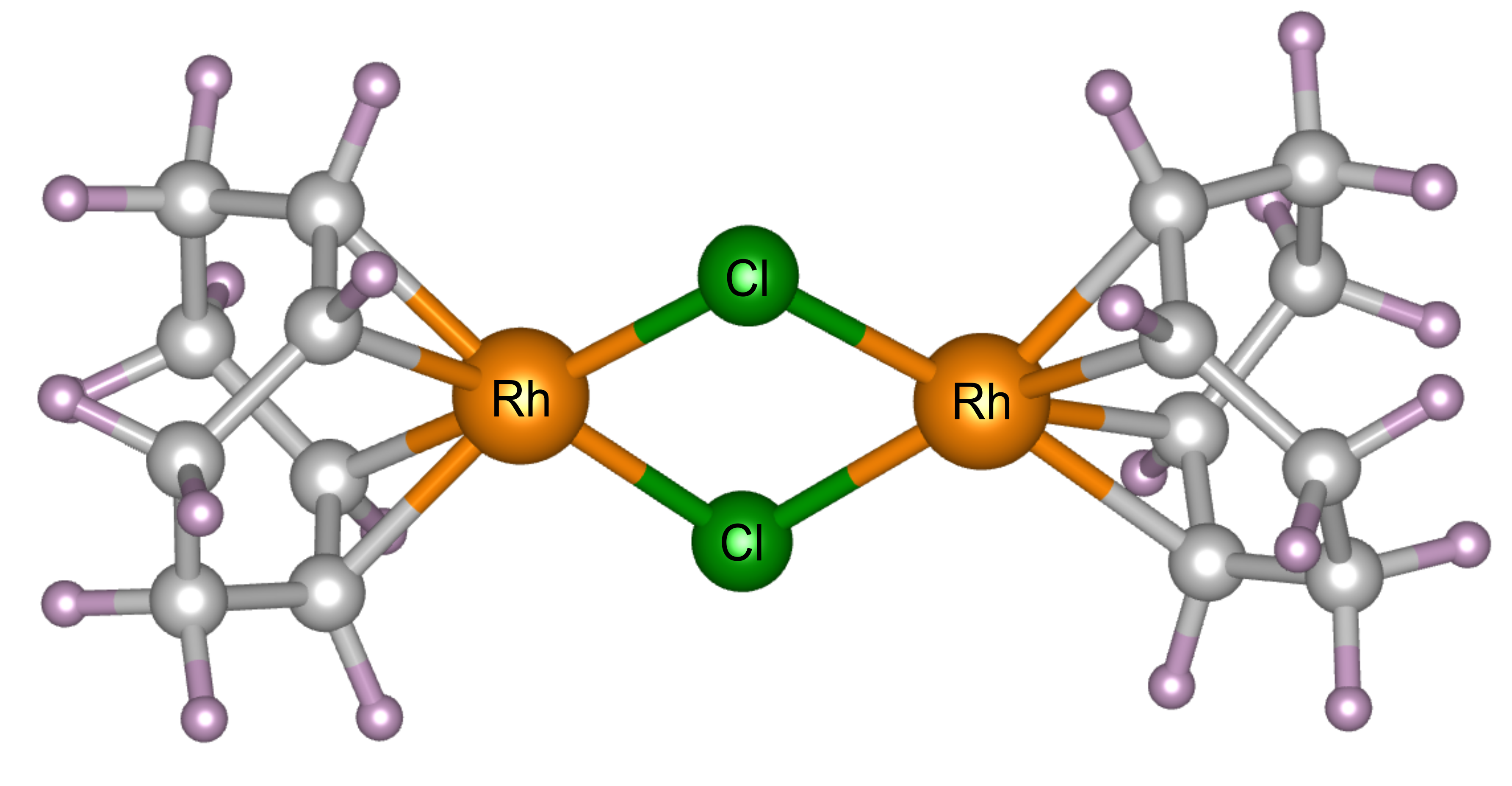}
\caption{The molecular unit of \ce{Rh(COD)Cl]2}, comprising of Rh (orange) and Cl (green) atoms, bonded to 1,5-cyclooctadiene, comprising of C (grey) and H (purple) atoms.}
\label{fig:Rh_str}
\end{figure}

This work highlights the necessity of applying dose-related MX experimental protocols in small molecule experiment planning, by exploring the collective effects on the sample using four different experimental arrangements at room temperature. A range of experiments were carried out at varying incident X-ray energies of 8, 15, 18 and 25~keV, which covered the photon energy range typically accessible at synchrotron powder diffraction beamlines. First, the strong, collective influence of four experimental arrangements on the global structure of \ce{[Rh(COD)Cl]2} is presented. The influence of the four incident energies on radiation damage and signal quality is discussed, identifying the optimum setting for the greatest diffracted signal per dose for \ce{[Rh(COD)Cl]2}. In addition, the applicability of dose limits (a form of quantification of X-ray hardness of a material and commonplace in MX) to organometallic systems, is discussed.

\section{Methods}
\subsection{Powder X-ray Diffraction}
Powder X-ray diffraction experiments were carried out at 300~K (controlled using a cryostream)~\cite{Cosier1986} at beamline I11 at Diamond Light Source, Didcot, UK. \ce{[Rh(COD)Cl]2} powder, acquired from Sigma Aldrich (ID: 227951) with 98\% purity and a molecular weight of 493.08~g/mol, was loaded into borosilicate capillaries with an outside diameter of 0.3~mm (and wall thickness of 0.01~mm) and placed in the path of the X-ray beam. The Rh complex crystallises in the monoclinic $P2_{1}/n$ space group. Its crystal structure has previously been determined by SCXRD.~\cite{Fernando2021} There is one crystallographically unique \ce{[Rh(COD)Cl]2} dimer in the asymmetric unit, each comprising 44 atoms, giving four molecules in the unit cell and a total of 176 atoms. The complex has a square-planar geometry and shares an edge in the Cl--Cl plane, see Figure~\ref{fig:Rh_str}.~\cite{Ibers1962, DeRidder1994} Four different experiments with incident photon energies typically accessible at diffraction beamlines were performed, namely 8, 15, 18, and 25~keV, corresponding to wavelengths of 1.549803(10), 0.826561(10), 0.688801(10) and 0.495937(10)~\AA, respectively. At each energy, 500 diffraction patterns were collected consecutively within a measurement window of two~hours, using the beamline's position sensitive detectors (PSD),~\cite{Thompson2011} and replicating the experimental procedure previously carried out by the authors.~\cite{Fernando2021}\par

Batch profile (Le Bail) and structural (Rietveld) refinements were conducted on all four datasets,~\cite{LeBail1988, Rietveld1967, Rietveld1969} using the TOPAS Academic~v7 analytical software, to quantify the structural changes to the Rh catalyst, observed at the different energies.~\cite{TOPAS7, Coelho2018} The background for all datasets was fitted using an eight order Chebyshev polynomial and the zero point error refined from a Si standard. A rigid body approach was used during the Rietveld refinements to model the complex structure of the COD ligands to limit the number of variable parameters in the refinement.~\cite{Scheringer1963} The atomic coordinates of the carbon atoms of the initial input file were defined using z-matrices, determined from the previously determined ``undamaged'' SCXRD structure solution of the Rh catalyst. The Rh and Cl positions were also defined using this previously determined structure, although these coordinates were allowed to refine freely throughout all 500 iterations. The rigid body model allowed the COD structure to rotate and translate as a entire body about the central Rh—Cl rhombus. Thermal parameters were allowed to refine during the Rietveld refinements of all datasets. \par

Refining the isotropic thermal displacement parameters, B\textsubscript{\textit{eq}}, with the only constraint being that the value should remain positive, led to unstable B\textsubscript{\textit{eq}} values for the C atoms. In addition, for the energies probed, the C B\textsubscript{\textit{eq}}s saw a sudden decline to zero during the batch refinement. Testing the various output B\textsubscript{\textit{eq}} values for C, as a starting point to the batch refinement, the final structure obtained from the refinements were observed to be insensitive to small discrepancies in B\textsubscript{\textit{eq}}. For this reason, and to avoid further complicating the refinement of such a complex system, the B\textsubscript{\textit{eq}} for C were fixed. 

\subsection{Estimation of X-ray dose}
The progression of damage within a system is expected to be proportional to the absorbed X-ray radiation dose. To better compare damage of the Rh catalyst across the beamline settings at varying photon energies (and associated fluxes) studied, dose was estimated using the RADDOSE-3D tool, which accounts for the beam properties, including energy, horizontal and vertical profile, and flux, as well as the material properties, such as the chemical composition.~\cite{Zeldin2013, Bury2018} The small molecule crystal addition to the calculations by Christensen~\textit{et al.}, was utilised, as well as the capillary model introduced by Brooks-Bartlett \textit{et al.}~\cite{Christensen2019, Brooks2017} Values of photon flux at each of the four energies at the beamline, were obtained by cross-calibrating current values collected using a Canberra PD 50-11-500aB Si PIN photodiode placed in-line as close as possible to the sample position ($-11.5$~mm), while a calibrated Canberra PD300-500CD photodiode was used for cross-calibration. This was carried out by recording the calibrated diode current at the sample position at the four energies and determining the corresponding photon flux as a function of a beam position monitor photocurrent (placed immediately before the sample position). The above steps were repeated, now with the in-line I11 photodiode, to determine the calibrated photon flux as a function of I11 diode current.~\cite{Owen2009} The Diffraction Weighted Dose (DWD) metric, output by RADDOSE-3D, was deemed the most applicable according to the experimental arrangement. A limitation of the dose estimations is the lack of real-time and accurate characterisation of beam profile and sizes measured at the different energies at the beamline, which is known to be crucial for dose determination. The value of beam size used in this study was obtained from beam characterisation undertaken by Thompson~\textit{et al.}~in 2009 at 15~keV, see Figure~2(a) and (b) in Ref.~\citenum{Thompson2009}. In addition to the limitation of using the beam shape and size at 15~keV for all energies studied here, since the beamline was last characterised, there is a possibility that the beam profile may have changed. This highlights the necessity of carrying out regular and systematic beam characterisation at synchrotron beamlines to enable inter-comparisons and quantative assessments of experiments. A full list of the RADDOSE-3D input parameters used in the DWD calculations are included in Table 1 of the Supplementary Information.\par

\section{Results \& Discussion}

\begin{figure}[ht]
\centering
\includegraphics[width=0.495\textwidth]{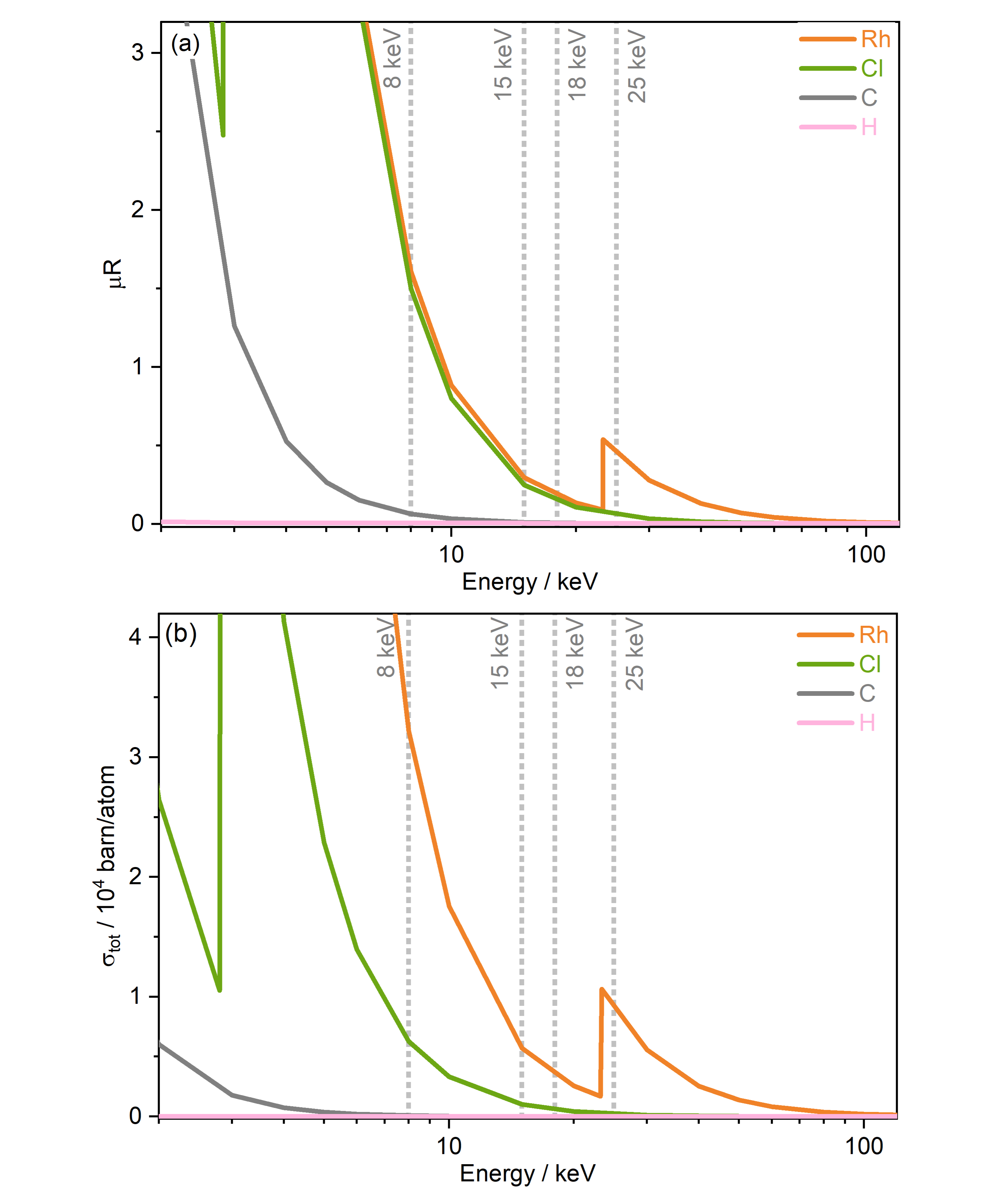}
\caption{Energy-dependence of (a) the absorption coefficients $\mu R$ obtained from WebAbsorb, Argonne National Laboratory,~\cite{Argonne2013} and (b) the total photoionisation cross sections $\sigma_{tot}$ from Scofield for elements present in \ce{[Rh(COD)Cl]2}.~\cite{Scofield_1973, Dig_Sco_2020}}
\label{fig:abs_phtn}
\end{figure}

Two important sample properties typically considered prior to conducting SMXRD and PXRD experiments are the total absorption coefficients and the total photoionisation cross section of the sample studied and their dependence on X-ray energy. Both quantities are well described by theoretical models, and Figure~\ref{fig:abs_phtn} shows both the absorption coefficients $\mu R$ extracted from the WebAbsorb utility and the total photoionisation cross sections $\sigma_{tot}$ from Scofield for the Rh complex.~\cite{Argonne2013, Scofield_1973, Dig_Sco_2020} For both quantities, the absorption edges of Rh and Cl are clearly visible. The 8~keV photon energy intersects the tails of the Rh L-edge and Cl K-edge, with 15 and 18~keV located towards the end of the tails. Meanwhile, 25~keV sits just above the Rh K-edge. The photoionisation cross sections show that considering only the incoming photon energy, the order of probability for ionisation events for this compound from greatest to lowest is 8, 25, 15, and finally 18~keV.\par

An understanding of the positions of element-specific absorption edges has allowed small molecule experimentalists to avoid ``problematic'' energy ranges in which the proportion of absorbed photons is high by varying the excitation energy. In addition to limiting absorption, energies are typically chosen with the aim of obtaining high resolutions, for instance to resolve molecular features such as bond lengths. In this study, we argue that considering only the energy dependence of absorption/photoionisation cross sections is insufficient, and other equally important energy-dependent experimental factors must be considered to determine ideal measurement conditions that minimise unwanted radiation effects.

The photon flux (in ph/s) across the X-ray energy range used in the present experiment is not constant, due to the necessary adjustments to the beam optics required to focus the different beams. Thus changing the photon energy leads to differences in photon flux and consequently, the dose absorbed by the sample. The DWD as a function of X-ray exposure time, determined from RADDOSE-3D for each of the four incident energies studied, is presented in Figure S1 of the Supplementary Information. The RADDOSE-3D input parameters used to estimate the absorbed dose are displayed in Table~S1. Table~\ref{tbl:flux_dose} contains the photon flux $\Phi$ at each of the four energies studied, along with the maximum dose values DWD\textsubscript{(\textit{t}=41.7 min)} corresponding to 41.7~min of total sample irradiation. These dose estimations enable a prediction to be made of the correlation of the extent of structural damage at the different photon energies used. At the 8~keV photon energy setup, for a given X-ray exposure time, the sample absorbs the greatest dose, relative to that at the other photon energies, by a substantial margin, due to the higher photon flux, absorption coefficients and photoionisation cross sections. In other words, at 8~keV, the dose rate is the highest. In addition, the elastic scattering cross section, which contributes to diffraction, is the highest at the lowest energy, 8~keV. At 15~keV the sample absorbs the next highest dose, followed by at 18~keV, and lastly at 25~keV. Whilst the absorption and photoionisation cross sections are higher at 25~keV than at 15~keV and 18~keV, the photon flux is significantly lower relative to the other three energies; of the order of 10$^{11}$~ph/s is available at beamline I11 at this energy. As a result, the dose absorbed by the sample is substantially lower at 25~keV, relative to the other photon energies studied. The amount of flux available at a given energy is dependent on the photon source of the beamline. At beamline I11, the photon source is a 22~mm period in-vacuum undulator (IVU).~\cite{Thompson2009} The energy of the photons provided by the undulator is dependent on the amplitude of the periodic magnetic field produced in the centre of the two parallel periodic magnet arrays of the undulator. The amplitude of this field is varied by changing the gap; a larger gap reduces the peak magnetic field of the undulator, which reduces both the photon flux and the wavelength of the emitted photons, i.e the photon energy increases with larger gap.~\cite{Thompson2009}\par

\begin{table}[ht]
\centering
\small
  \caption{\ The photon flux $\Phi$ at the four photon energies $h\nu$ obtained from a calibrated photodiode, along with the corresponding maximum DWD at the end of the 41.7~min X-ray irradiation time DWD\textsubscript{(\textit{t}=41.7~min)} and the fraction, $Q$, of the maximum DWD relative to the 8~keV value. The elastic scattering coefficient, $\sigma$, obtained from RADDOSE-3D, is also presented.}
  \label{tbl:flux_dose}
  \begin{tabular*}{0.48\textwidth}{@{\extracolsep{\fill}}lllll}
    \hline
    $h\nu$ / keV & $\Phi$ / ph/s & DWD\textsubscript{(\textit{t}=41.7~min)} / MGy & $Q$ & $\sigma$ / 10$^{-4}$$\mu$m$^{-1}$\\
    \hline
    8 & 5.7 $\times$ 10$^{12}$ & 53 & 1 & 2.97\\
    15 & 2.4 $\times$ 10$^{12}$ & 12 & 0.2 & 1.48\\
    18 & 1.4 $\times$ 10$^{12}$ & 6 & 0.1 & 1.17\\
    25 & 2.2 $\times$ 10$^{11}$ & 1 & 0.02 & 0.74\\
    \hline
  \end{tabular*}
\end{table}

\begin{figure}[ht]
\centering
\includegraphics[width=0.495\textwidth]{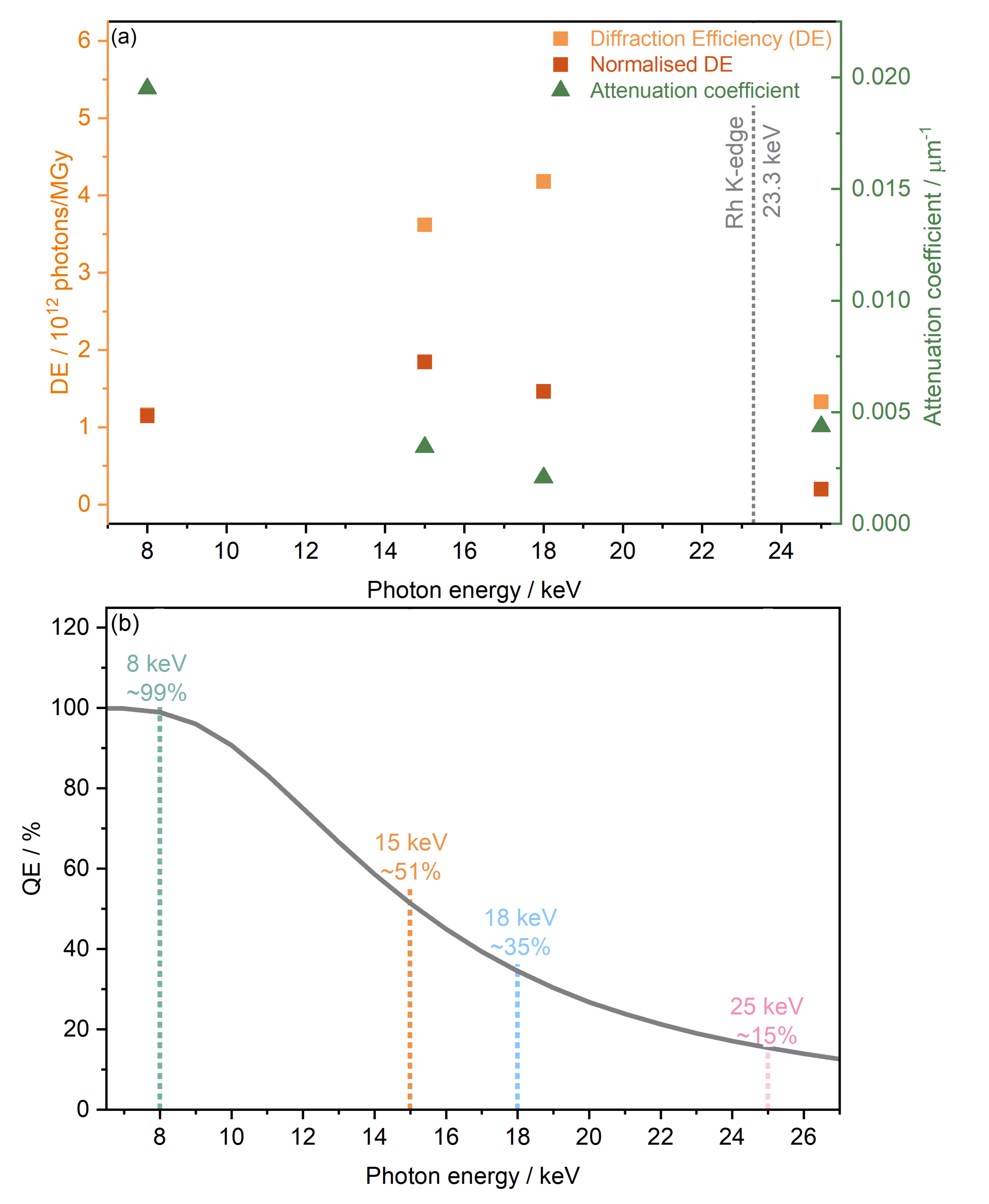}
\caption{The diffraction and quantum efficiencies of the Rh complex and the detector, respectively. (a) The estimated diffraction efficiency (DE) of \ce{[Rh(COD)Cl]2} obtained from RADDOSE-3D calculations and diffraction efficiencies normalised to the theoretical quantum efficiency of the detector (see subfigure b), at 8, 15, 18, and 25~keV, plotted alongside the corresponding attenuation coefficients. (b) The theoretical quantum efficiency of the Mythen-II DECTRIS detector, used at beamline I11, Diamond Light Source, determined from the attenuation percentage of 300~$\mu$m thick Si, as a function of photon energy.} 
\label{fig:DE}
\end{figure}

Based on dose calculations, another factor often considered in MX studies when determining ideal photon energy is the diffraction efficiency (DE). This is defined as the total number of elastically scattered photons per X-ray dose and provides theoretical information on which photon energy will have the highest signal to dose ratio, and thus minimises radiation-induced effects. The diffraction efficiency of \ce{[Rh(COD)Cl]2} which was obtained from RADDOSE-3D calculations at 8, 15, 18, and 25~keV, is presented in Figure~\ref{fig:DE}(a). Considering the DE alone, it is clear to see that of the energies studied, the maximum DE occurs at 18~keV, just before the Rh K-edge, seemingly making it the ideal energy at which to probe. This is closely followed by 15~keV. The lowest DEs are found at 25~keV and 8~keV. However, as outlined in the work of Dickerson~\textit{et al.}, the DE is also affected by the quantum efficiency of the photon detector used.\cite{Dickerson2019} As mentioned in the Methods section, for high-throughput powder diffraction experiments, the Mythen-II PSD detectors of beamline I11, were used. These comprise Si modules of 300~$\mu$m thickness and a quantum efficiency of $\approx$100\% at 7~keV and ~15\% at 25~keV.~\cite{Thompson2011} Figure \ref{fig:DE}(b) shows the drop in quantum efficiency which corresponds to the attenuation percentage of the X-ray beam through Si of 300~$\mu$m thickness, assuming a detector threshold of 50\%.

The quantum efficiency is seen to drop from close to 99\% at 8~keV to $\sim$53\% at 15~keV, then $\sim$35\% at 18~keV and finally $\sim$15\% at 25~keV. Considering both the diffraction and quantum efficiencies at the four energies studied, at a photon energy of 15~keV the greatest signal to dose, \textit{i.e.} maximising the intensity whilst minimising radiation damage, is expected. 

In the above discussions, theoretical approaches have been used to predict and compare the overall radiation sensitivity of \ce{[Rh(COD)Cl]2} at the different energies studied. Next, the impact on the crystal structure is explored, as a function of the collective differences in experimental settings at the various energies. These collective differences are a result of the associated changes in flux, dose rate, dose \textit{etc}. when changing the X-ray energy. Estimating the theoretical dose, and diffraction efficiency, allows for the prediction of behaviour that is subsequently observed in experiment. Given that the dose at the experimental settings achieved at 8~keV is the largest relative to the other energy setups, it is expected that the structure will be perturbed the most at this setting during PXRD measurements. Conversely, at 25~keV, the smallest structural distortion is expected, according to the dose. However, taking into consideration the diffraction and quantum efficiencies, it can be seen that at 25~keV, resolution (signal-to-noise) will be very poor as the elastic scattering cross section and photon flux is at a minimum. At 18~keV, it is predicted that structural changes would again be minimal, although slightly greater than at 25~keV. In addition, it is clear from the diffraction (and quantum) efficiency that in this regime, resolution is maximised.\par Estimations of PXRD resolution, $q$, determined using the 2$\sigma$ procedure outlined in the Supplementary Information, agree with these findings, see Table S2. When ordering the first, minimum dose diffraction patterns by resolution, the 15~keV dataset had the highest resolution, followed by the 18~keV, 8~keV and finally 25~keV dataset. This corresponds to resolutions (d-spacings in ~\AA) of 6.20~\AA$^{-1}$ (1.01), 5.58~\AA$^{-1}$ (1.12), 4.06~\AA$^{-1}$ (1.55) and 3.29~\AA$^{-1}$ (1.91), respectively. Based on these values of d-spacing, structures from the 8~keV and 25~keV datasets may be less reliable, relative to the 15~keV and 18~keV datasets.   

\subsection{Changes to the unit cell}
Following the methodology developed in previous work by the authors,~\cite{Fernando2021} and in order to elucidate changes to the diffraction pattern with increasing cumulative X-ray dose, Le Bail refinements were carried out to determine changes to lineshape (FWHM), peak position (2$\theta$), and peak intensity (integrated area). These changes, for five low-angle Bragg reflections, at the four energies, are presented in Figure S2 of the Supplementary Information. It is evident that the peak changes follow the order of X-ray dose, as expected, with the highest dose, 8~keV dataset showing the most severe decay in peak intensity, increase in full width at half maximum, FWHM, and peak shift, followed by 15~keV, then 18~keV and finally, 25~keV which shows the smallest relative differences, of the experimental settings studied. The 8~keV Le Bail refinement, see Figures S2(a)-(c), underlines the anisotropic nature of the changes. The five reflections probed, underwent a 42\% to 55\% loss in peak intensity. In addition, the FWHM increase at different rates in each crystallographic direction, with the (011) reflection increasing in width by approximately 40\%, compared to the almost 110\% increase in the (130) reflection. It should be noted that the change in FWHM increases linearly across all directions, until approximately 40~min of irradiation, equivalent to a dose of 18~MGy at the 8~keV setup, plateauing to reach a maximum value after 26~MGy, equivalent to 60~min of irradiation. \par

This trend is also reflected in the change in peak positions for the 8~keV dataset, which shows a decrease and subsequent plateauing of the rate of peak shift after 60~min of X-ray exposure, across the different crystallographic directions studied. For example, the (011) peak undergoes a major shift towards lower 2$\theta$ values of $-1.3$\%, whereas the (110) sees a minimal $-0.3$\% shift. Le Bail refinements of the 15~keV datasets show less severe, though notable, intensity decays, FWHM increases, and a maximum (011) peak shift of $-0.5$\%, see Figures S2(d)-(f). Changes in the 18~keV dataset are even smaller, with the peak positions only undergoing a notable maximum shift of approximately $-0.2$\%, see Figures S2(g)-(i). Finally, the peak shape, intensity and peak positions of the 25~keV dataset over the same two hour irradiation period, see Figures S2(j)-(l), are too subtle to quantitatively extract, as would be expected from the low X-ray dose at the experimental settings corresponding to this energy.\par

\begin{figure*}[ht]
\centering
\includegraphics[width=\textwidth]{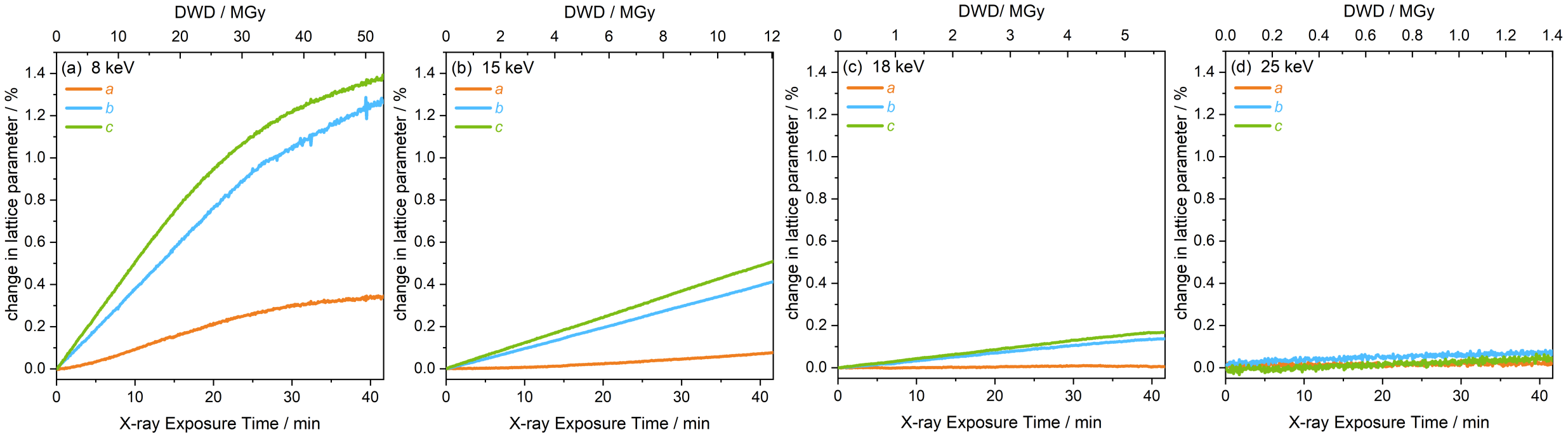}
\caption{Change in the lattice parameters $a$, $b$ and $c$ of \ce{[Rh(COD)Cl]2} as a function of diffraction-weighted dose (DWD) and X-ray exposure time at photon energy setups of (a) 8~keV, (b) 15~keV, (c) 18~keV, and (d) 25~keV.}
\label{fig:lattice_param}
\end{figure*}

The extent of peak shifts across all four energies is reflected in the change in the unit cell lattice parameters, also determined by full profile batch Le Bail refinements, shown in Figure~\ref{fig:lattice_param}. The $b$ and $c$ lattice parameters clearly increase substantially more than the $a$ lattice parameter over the same dose and time scales. The discontinuity in peak shape in the 8~keV data, after 26~MGy (\textit{t}=21~min) is also evident in the change to lattice parameters in Figure~\ref{fig:lattice_param}(a). After this discontinuity, the onset of a plateau is observed across $a$, $b$ and $c$ at 8~keV. Relative to the dose at the 8~keV setting, a significantly lower dose is absorbed by the sample at experimental settings corresponding to 15, 18 and 25~keV. This results in the increase in lattice parameters remaining almost linear throughout the two hour period and a plateau is never reached.\par 

\begin{figure}[ht]
\centering
\includegraphics[width=0.495\textwidth]{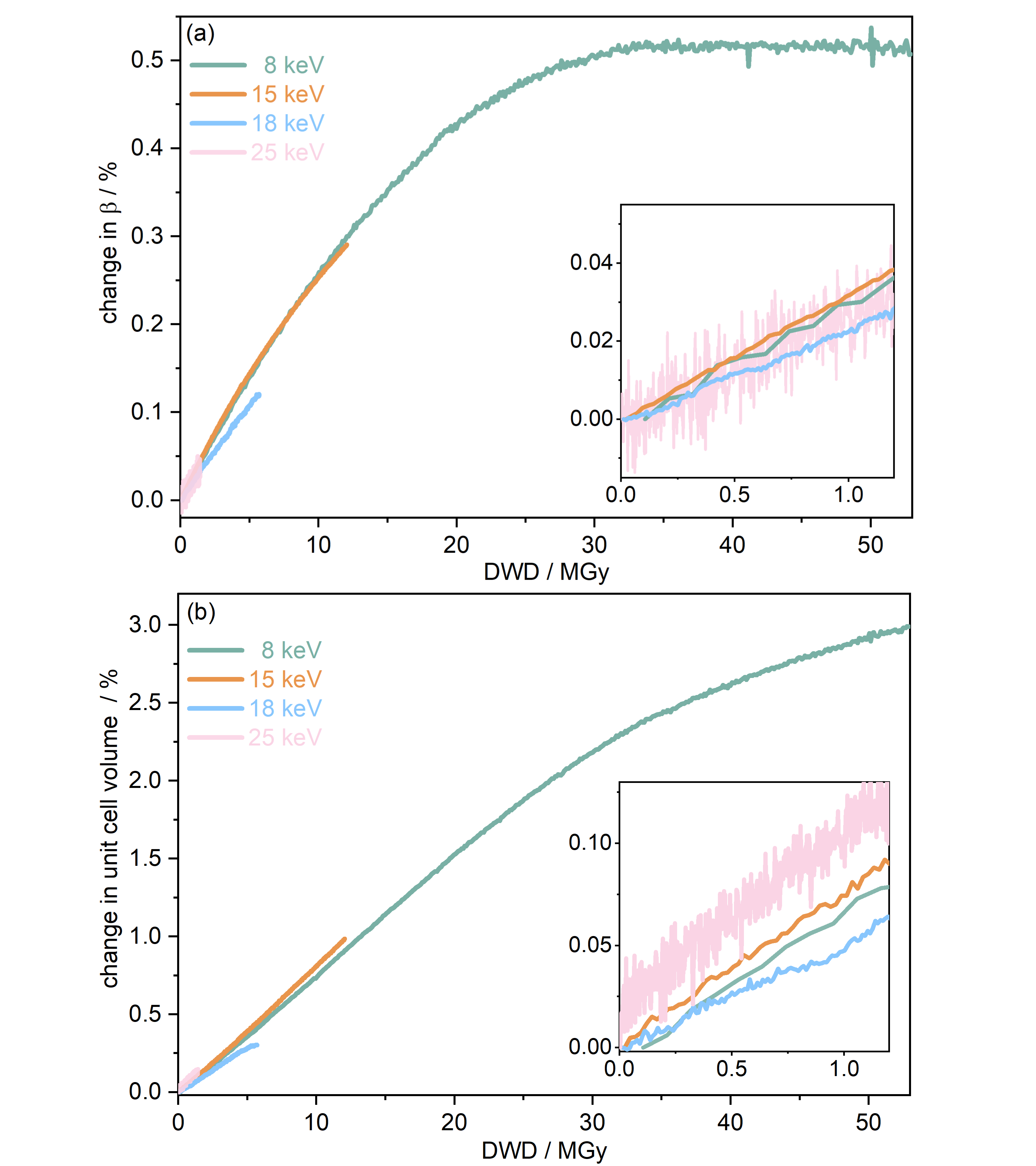}
\caption{The percentage change in (a) angle $\beta$ and (b) unit cell volume $V$ as a function of diffraction-weighted dose (DWD) at photon energies $h\nu$ of 8, 15, 18, and 25~keV. The insets show the low dose range up to 1.2~MGy, including labels for the corresponding X-ray exposure time. Note the wide range (order of magnitude difference) in the maximum DWD at the various incident energies.}
\label{fig:vol}
\end{figure}

To further understand the overall changes to the unit cell with increasing dose at the different photon energies, trends in the $\beta$ angle and unit cell volume were also explored and are shown in Figure~\ref{fig:vol}. As shown in Figure~\ref{fig:vol}(a) at 8~keV, $\beta$ stabilises after a 0.5\% increase relative to its original value after receiving a dose of approximately 35~MGy, and remains constant for the remainder of the X-ray exposure. The global $\beta$ angle of the monoclinic Rh complex first undergoes an almost linear rate of expansion of $0.03^{\circ}$/MGy until approximately 13~MGy, from which point the rate of expansion gradually decreases until it reaches an upper limit of radiation-induced expansion at a dose of approximately 35~MGy, at which point $\beta$ is 92.37$^{\circ}$ (a 0.5\% increase from the starting angle of 91.89$^{\circ}$). From the point of constant $\beta$, the unit cell volume continues to increase solely through changes in $a$, $b$ and $c$, but the rate of expansion slows down significantly from 0.004~\AA$^{3}$/MGy to 0.002~\AA$^{3}$/MGy. Considering the relatively low doses of the 15~keV to 25~keV experiments, the lattice parameters, and thus the unit cell volume, continually increase across the two-hour time frame of the experiments without reaching a plateau, as evident in Figure~\ref{fig:vol}(b). At 15~keV the Rh complex experiences a linear unit cell volume increase of 0.98\%, from 1688.7~\AA$^{3}$ to 1705.4~\AA$^{3}$, at a rate of 0.008~\AA$^{3}$/MGy and at 18~keV, it expands by 0.30\% from 1679.7~\AA$^{3}$ to 1684.7~\AA$^{3}$ at a rate of 0.005~\AA$^{3}$/MGy. Due to the extremely low flux achieved at 25~keV, the signal resolution of the diffraction patterns is severely compromised, as evidenced by the high level noise in the profile refinements in Figure~S2 of the Supplementary Information. This is in agreement with the lowest predicted diffraction and detector efficiency at 25~keV determined using RADDOSE-3D in Figure~\ref{fig:DE}(a). For this reason, it is not possible to quantify the changes in the 25~keV dataset. The changes in $\beta$ and $V$ as a function of dose, $\frac{d\beta}{dD}$ and $\frac{dV}{dD}$, are comparable across the different energies at low dose values, see insets in Figure~\ref{fig:vol} and Table~\ref{tbl:dose_rate}. The differences in these unit cell changes become slightly more apparent at higher dose. At a given dose, the 15~keV dataset appears to undergo the greatest change in $\beta$ and volume (excluding the 25~keV dataset), closely followed by 8~keV and lastly 18~keV, although these differences are not significant at low doses of less than 1.2~MGy. These minor discrepancies at higher dose could be understood through the combined effects of the difference in dose rates and resolution at the different energies, as well as the lack of accurate knowledge of beam profiles.\par 

Southworth-Davies~\textit{et al.}~observed that at room temperature on a home X-ray source in MX, the $D_{0.5}$ value, \textit{i.e.} the dose at which half of the diffraction intensity is preserved, increases linearly with dose rate, in Gy s$^{-1}$.\cite{Davies2007}. Expanding on this study, Owen~\textit{et al.}~showed a dose-rate effect at room temperature at an undulator beamline, with crystal lifetimes increasing as a function of dose rate.~\cite{Owen2012} In order to determine conclusively the presence of dose-rate dependence, the photon flux across the different energies should be attenuated such that they are approximately equal in value. In reality, and in part due to physical beamline limitations (\textit{i.e.} type of undulator), this is not a trivial pursuit.

\begin{table}[ht]
\centering
\small
  \caption{\ Dose rates associated with the photon energies $h\nu$ studied, along with the corresponding change per unit dose, $D$, in angle $\beta$, $\frac{d\beta}{dD}$, and unit cell volume $V$, $\frac{dV}{dD}$.}
  \label{tbl:dose_rate}
  \begin{tabular*}{0.48\textwidth}{@{\extracolsep{\fill}}lrrr}
    \hline
    $h\nu$ / keV & Dose rate / MGy s$^{-1}$ & $\frac{d\beta}{dD}$ / $^{\circ}$MGy$^{-1}$ & $\frac{dV}{dD}$ / \AA$^{3}$MGy$^{-1}$\\
    \hline
    8  & 1.3 & 0.0267 & 0.0749\\
    15  & 0.3 & 0.0299  & 0.0776 \\
    18  & 0.1 & 0.0219 & 0.0512\\
    25  & 0.03 & 0.0286 & 0.0878 \\
    \hline
  \end{tabular*}
\end{table}

\subsection{Changes to the molecular unit}
Le Bail refinements of the PXRD data provide insights into the average unit cell changes occurring in the sample irradiated at the different photon energies. However, in order to further probe the energy and related experimental parameter-dependent structural changes experienced by each molecular unit comprising the unit cell, batch Rietveld refinements of all 500 diffraction patterns across the 2-hour time frame of the experiments were carried out. \par

Previous iterative X-ray Photoelectron Spectroscopy (XPS) studies of this system showed that the carbon atoms comprising the COD ligands undergo no substantial radiation-induced chemical environment change.~\cite{Fernando2021} The rigid body model for the COD proved incredibly useful for the comparative batch refinements across the four energies, as it enabled the carbon atoms to preserve their chemical environment throughout the sample irradiation and shift relative to the central M--Cl unit. This method also allowed the extraction of the changes in the central Rh--Cl core structure more robustly, which is known to be sensitive to X-ray radiation.~\cite{Fernando2021} \par

\begin{figure*}[ht]
\centering
\includegraphics[width=\textwidth]{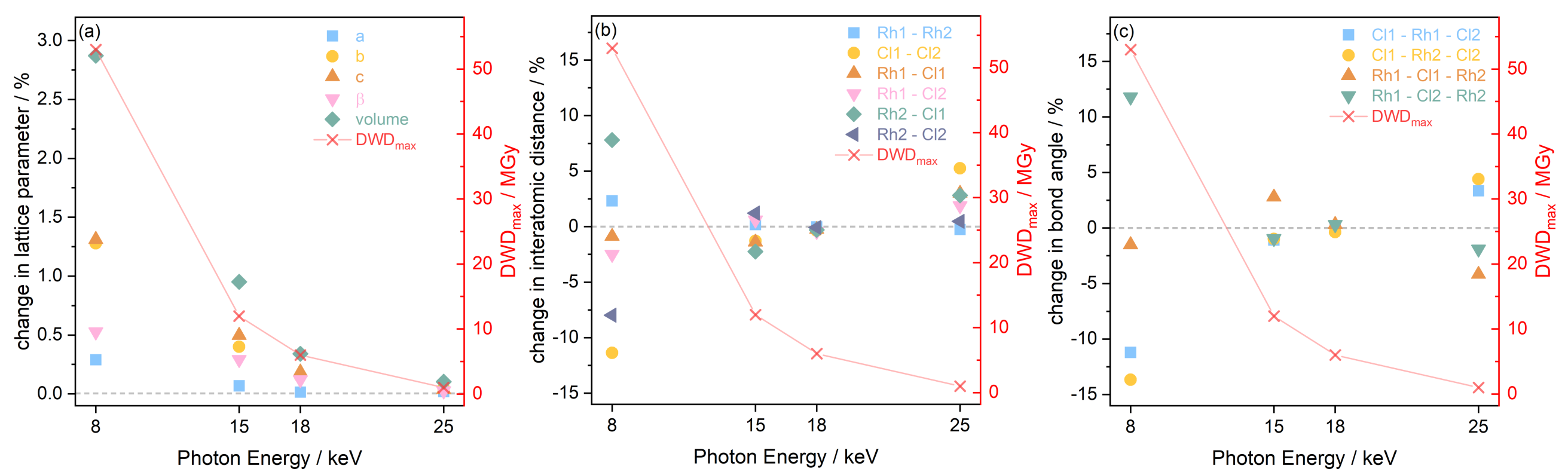}
\caption{Structural changes observed in \ce{[Rh(COD)Cl]2} at the four photon energies $h\nu$ (8, 15, 18 and 25~keV) with respect to the maxmimum diffraction-weighted dose (DWD\textsubscript{max}) after 41.7~min of X-ray exposure, including the percentage changes in (a) the lattice parameters $a$, $b$ and $c$, (b) the interatomic distances, and (c) the bond angles.}
\label{fig:structure}
\end{figure*}

The results extracted from the structure refinements, namely the change in the lattice parameters, the interatomic distances, and bond angles, are presented in Figure~\ref{fig:structure} as a function of photon energy, from the first, minimum dose diffraction pattern to the maximum dose diffraction pattern (after two hours of irradiation). It is evident in Figure~\ref{fig:structure}(a) that the trend of greatest lattice parameter expansion at the experimental setting associated with a photon energy of 8~keV, is consistent with the Le Bail refinement discussed above. In addition, the extent of expansion reduces at increasing photon energy, and hence, decreasing dose, with the 25~keV sample experiencing almost no notable change. The increase in lattice parameters with X-ray dose has been previously reported on this compound at room temperature~\cite{Fernando2021} and has also been observed in proteins in MX at 100~K by Murray~\textit{et al.}~and Ravelli~\textit{et al.}.~\cite{Murray2002, Ravelli2002}\par 

It should be noted that the changes presented in the interatomic distances and bond angles of atoms present in the average molecular unit are not only a direct consequence of radiation-induced sample change, but are also convoluted by the resolution of the minimum and maximum dose diffraction patterns. At the 8~keV photon energy, at an absorbed dose of 53~MGy (\textit{t}=41.7~min), the atomic positions are significantly perturbed. This is evidenced by the widespread changes to the interatomic distances and bond angles, see Figures~\ref{fig:structure}(b) and (c). This is particularly clear for Cl1--Cl2, which sees an 11\% decrease in distance, whilst the Rh2--Cl1 distance undergoes an expansion of approximately 8\%. At 8~keV, the Rh1--Cl2--Rh2 angle increases by 11\%, whilst the Cl1--Rh2--Cl2 angle decreases by 14\%, as expected from earlier dose calculations, since the absorbed dose decreases for the datasets collected at higher energy, and the extent to which interatomic distances change becomes gradually smaller. At 15~keV, the changes to the atomic distances and bond angles are less than 3\% from the starting structure. At 18~keV, which has a maximum dose of 6~MGy, there are no notable changes in either the interatomic distances or the bond angles, between the first and last measurement points.\par

The high dose absorbed by the Rh complex during data collection at 8~keV leads to an anisotropic distortion of the central M--Cl structure, which is in agreement with previous observations by the present authors on the related \ce{[Ir(COD)Cl]2} complex.~\cite{Fernando2021} The Cl atoms appear to be the more mobile species compared to Rh, which is also reflected in the rising Cl isotropic atomic displacement parameters with increasing cumulative dose, see Figure S3 in the Supplementary Information. Figure~\ref{fig:8kev_structure} illustrates this by focusing on the Rh--Cl motif, where radiation-induced change manifests as a bending of the initially planar Rh--Cl structure about the Rh--Rh axes such that the Cl--Cl distances are reduced. The authors' previous XPS analysis of this system showed that the central Rh--Cl structure experiences a severe loss of Cl as a result of X-ray irradiation. We were unable to attribute the Cl loss in the Rh complex to any structural observations from the complementary diffraction studies conducted at 18~keV, since the changes were too subtle to observe in \ce{[Rh(COD)Cl]2} due to the low absorbed dose at this energy. However, since the present study also includes data collected using the high flux and dose available at 8~keV, the global structural changes to the Rh--Cl structure are clearly evident in the structure from that dataset. As a result, the notable global Cl displacement relative to its original, minimum dose position, and relative to the approximately stable Rh atoms, can be correlated with the local movement of the Cl away from the central Rh--Cl core, observed in XPS.~\cite{Fernando2021}\par 

\begin{figure}[ht]
\centering
\includegraphics[width=0.475\textwidth]{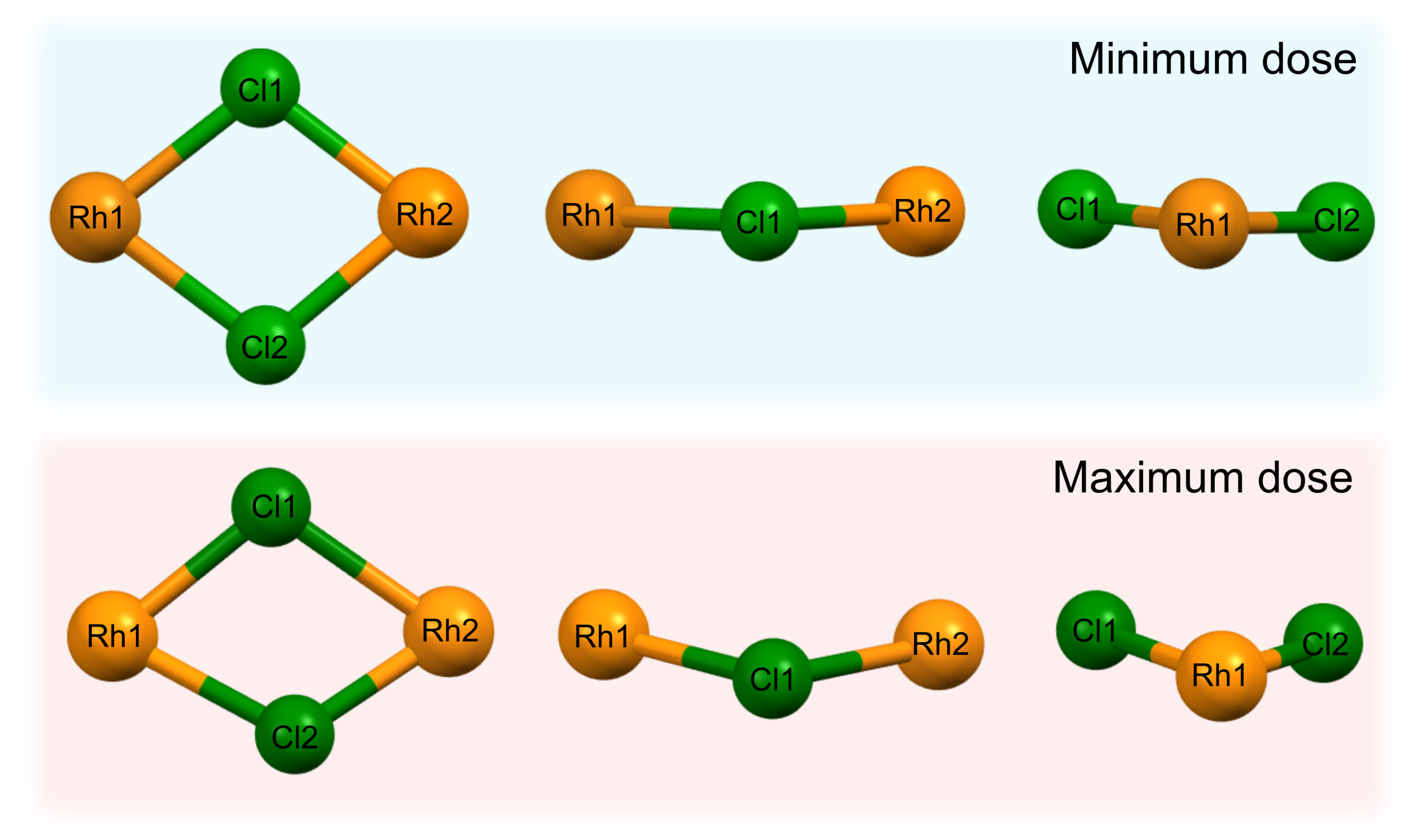}
\caption{The central Rh--Cl motif of \ce{[Rh(COD)Cl]2} obtained from refinement of the dataset collected at 8~keV photon energy, overlooking the Rh--Cl plane, and the side profiles, depicting the change in the structure's atomic arrangement as a result of X-ray irradiation. The top row is the minimum dose structure (0.1~MGy, \textit{t}=5~s) and the bottom row is the maximum dose (53~MGy, \textit{t}=41.7~min) structure.}
\label{fig:8kev_structure}
\end{figure}

From Le Bail and Rietveld refinements, it is clear that the 25~keV dataset does not follow the expected trend. Whilst the dose is the lowest at 25~keV, the achieved diffraction data quality, due to the very low photon flux, and low elastic cross section is too poor to obtain reliable results. The noise levels are too significant to determine the structure of the Rh complex accurately. Given that the interatomic distances and bond angles determined by Rietveld refinements are sensitive to data resolution, it is no surprise to observe a spreading of the 25~keV data points shown in Figures~\ref{fig:structure}(b) and (c), reflecting the uncertainty at this energy. Despite the low dose achieved at 25~keV, the photon flux and elastic cross sections are too poor to achieve adequate signal-to-noise, in reasonable time. A preferable energy setting would be one in which the diffraction efficiency is maximised, but this experiment clearly demonstrates that it is also imperative to consider whether at that energy, the necessary resolution can be achieved in a reasonable time. In other words, there is a trade-off between limiting dose and achieving the necessary resolution to conduct the required data analysis.

\subsection{Tolerable Dose Limit Estimations}

In order to quantify the difference in diffraction data quality over the total measurement duration between the four photon energies, the total diffraction intensity $I_{tot}$, analogous to the dose limit determination in single crystal X-ray diffraction, was obtained, see Table \ref{tbl:intensity_loss}. This was extracted from the datasets by integrating the powder diffraction pattern at minimum dose (\textit{t}=5~s) and maximum dose (\textit{t}=41.7~min), across a specified angular range of 5-60$^{\circ}$ comprising all resolvable peaks as tabulated in Table S2 of the Supplementary Information. Despite the flux being the greatest at 8~keV, the total intensity is considerably low. This can be explained by the high absorption at this energy as it is situated on the tail of both the Rh L- and Cl K- edges. The proportion of total diffraction intensity remaining after 41.7~min sample irradiation is also presented. The 8~keV dataset is the only one to lose more than half the initial total diffraction intensity. The 15, 18 and 25~keV datasets preserve more than 89\% of the initial diffraction intensity after the total measurement time.\par 

\begin{table}[ht]
\centering
\small
  \caption{\ Total diffraction intensities $I_{tot}$ for photon energies $h\nu$ of 8, 15, 18, and 25~keV, at the minimum (\textit{t}=5~s) and maximum measurement dose (\textit{t}=41.7~min), along with the remaining proportion of intensity $x$ after the 41.7~min measurement period, where $x=I_{tot}(41.7~\textrm{min})/I_{tot}(5~\textrm{s})$.}
  \label{tbl:intensity_loss}
  \begin{tabular*}{0.48\textwidth}{@{\extracolsep{\fill}}rrrr}
    \hline
    $h\nu$ / keV & $I_{tot}$ ($t$=5~s) & $I_{tot}$ (\textit{t}=41.7~min) & $x$\\
    \hline
    8 & 22383 & 7467 & 0.33 \\
    15 & 76258 & 68120 & 0.89 \\
    18 & 30888 & 28476 & 0.92 \\
    25 & 776 & 763 & 0.98 \\
    \hline
  \end{tabular*}
\end{table}

The lifetime of a crystal under X-ray irradiation in MX experiments is typically quantified by dose limits, based on a defined point at which important biological information is deemed to be compromised. In the following discussion, three different definitions of dose limits cited in literature for protein crystals are explored and applied to the Rh complex. Table \ref{tbl:dose_limit} summarises the various dose limits calculated.\par

The Henderson-type dose limit is defined by the ``half-damage'' dose of the total diffraction intensity $D_{0.5}$ in a typical single crystal diffraction experiment at cryotemperatures. Here, only the high photon flux at 8~keV enables such a dose limit to be determined, with the Henderson-type dose limit calculated to be 39.6~MGy for room temperature powder data collection. However, based on the observations reported in this study, it is clear that even at this dose, substantial structural change has already occurred in the form of up to 2.5\% unit cell expansion, and movement in atomic positions. This suggests that $D_{0.5}$ is therefore not suitable for use in powder diffraction studies of organometallic crystals. In addition to the ``Henderson-type'' limit, Owen~\textit{et al.}~suggested another dose limit defined as the dose at which 70\% of the total diffraction intensity is retained in MX at 100~K.~\cite{Owen2006} Here, the experimental dose limit of $D_{0.7}$ is calculated to be at 26.1~MGy for the 8~keV dataset. Again, at this dose, although reduced compared with the $D_{0.5}$ intensity, significant structural distortion is observed, with an approximately 2\% increase in unit cell volume.\par

As both the ``Henderson'' and ``experimental dose'' (Owen) limits  are not suitable for this organometallic system at room temperature, an alternative approach of applying a $D_{0.9}$ dose limit is considered, defined as the dose at which 90\% of the initial diffraction intensity is retained. Room temperature MX dose limits (specifically dose rate effects) were investigated by De La Mora~\textit{et al.}. The authors proposed a 0.38~MGy dose limit but stressed that this was true only for hen egg-white lysozyme (HEWL) crystals at 2~\AA, and is therefore inapplicable to the organometallic system studied here.~\cite{Mora2020} The $D_{0.9}$ definition of dose limit has not, to the best of our knowledge, been previously applied to crystallographic experiments. However, a $D_{0.9}$ dose limit has been reported for a few X-ray absorption spectroscopy experiments of metalloproteins, defined by the upper dose at which 90\% of the intrinsic metal oxidation state is preserved.~\cite{Kubin2018} Based on our previous study including XPS of the \ce{[Rh(COD)Cl]2} system this seems to be a justifiable alternative. At 8 and 15~keV the $D_{0.9}$ limit is determined to be 11.9~MGy (corresponding to approximately 27~min of X-ray exposure and estimated resolution of $2.24\pm0.01$\AA$^{-1}$) and 11.8~MGy (equivalent to 40.6~min and resolution of $5.53\pm0.06$\AA$^{-1}$), respectively. This result is as expected, giving remarkable agreement, since by its very definition dose limit should be independent of incident X-ray energy.\par

At the $D_{0.9}$ limit, there is a minimal increase in lattice parameters and insignificant change in the planar Rh-Cl central core. Considering the limited structural change, this definition of a tolerable dose limit is more favourable to \ce{[Rh(COD)Cl]2} and possibly other organometallic crystals irradiated by X-rays during powder diffraction experiments. Determining $D_{0.9}$ could be important when systematically comparing the X-ray radiation hardness in related systems using precisely the same experimental parameters, including energy, flux, dose rate \textit{etc}. However, it is important to note that the reduction in crystallographic quality crucially does not account for any local chemical environment change, such as photoreduction processes in the irradiated sample. Understanding the extent of metal reduction during irradiation in an organometallic catalyst is vital. From previously published X-ray photoelectron spectroscopy data on \ce{[Rh(COD)Cl]2}, the dose at which 90\% of the Rh(I) state is preserved at room temperature, $D_{0.9}$, is determined to be approximately 7~MGy.~\cite{Fernando2021} Attempts to correlate structural with chemical damage at these XPS and XRD dose limits should be done cautiously given the varying experimental arrangements. PXRD is typically conducted in ambient conditions, in contrast to XPS which is carried out in an ultrahigh vacuum (UHV) environment. The specific local change, in the form of metal photoreduction has a much earlier onset in the damage process than the global structural change. At the PXRD $D_{0.9}$ limit of 11.8~MGy, despite the limited global damage to the unit cell, the Rh complex will have experienced substantial specific local damage in the form of Rh(I) reduction and Rh-Cl coordination changes. It is therefore important to consider both the crystallographic and spectroscopic $D_{0.9}$ limits to describe both the global structural and local chemical changes when discussing X-ray sensitivity of organometallic systems.\par

\begin{table}[ht]
\centering
\small
  \caption{\ Range of dose limits $D\textsubscript{x}$ of \ce{[Rh(COD)Cl]2} determined from powder diffraction datasets collected at photon energies $h\nu$ of 8 and 15~keV, including Henderson $D_{0.5}$, Owen $D_{0.7}$ and organometallic specific $D_{0.9}$. The maximum dose after 41.7~min of irradiation is 53~MGy and 12~MGy at the 8~keV and 15~keV energy settings, respectively.}
  \label{tbl:dose_limit}
  \begin{tabular*}{0.48\textwidth}{@{\extracolsep{\fill}}rrr}
    \hline
    $D\textsubscript{x}$ & $h\nu$ / keV & Dose limit / MGy\\
    \hline
    $D_{0.5}$ & 8 & $39.6~\pm~0.1$\\
    $D_{0.7}$ & 8  & $26.1~\pm~0.1$\\
    $D_{0.9}$ & 8 & $11.9~\pm~0.1$\\
    $D_{0.9}$ & 15 & $11.8~\pm~0.1$\\
    \hline
  \end{tabular*}
\end{table}

\section{Conclusions}
In this study, X-ray induced structural changes in the \ce{[Rh(COD)Cl]2} catalyst during room temperature PXRD experiments at four different X-ray experimental settings are explored. The severe structural distortion of the central Rh--Cl structure at 8~keV and up to 15\% change in bond angles, is attributed to the intense photon flux and large absorption and photoionisation cross sections, and therefore, radiation dose, associated with the experiment at this energy. The dose values at energies between 15 and 25~keV are an order of magnitude smaller than at 8~keV, resulting in more subtle structural changes. However, integrated intensities are lower at 8~keV than at 15~keV and 18~keV, see Table~\ref{tbl:intensity_loss}, and previously discussed estimations of resolution show that the 15~keV setting has the highest detectable resolution, followed by 18~keV, 8~keV and finally 25~keV, see Table~S2. This means that the expected changes to structure at 8~keV may not be detected from the diffraction patterns, especially as the 8~keV setting sees the largest drop in resolution of $-61\%$ after 41.7~min of irradiation. The significant movement of Cl atoms relative to the Rh atoms at 8~keV, can be correlated to the previously reported local loss of Cl during X-ray irradiation in photoelectron spectroscopy experiments. From the 8~keV dataset, insights into the onset of an upper limit of X-ray damage, where the unit cell changes appear to stabilise after approximately 35~MGy of irradiation, are obtained.\par

Although there were physical beamline limitations to determining a true energy-dependence of damage during this experiment, going forward, to better determine the optimal energy, the experiment should be setup such that equal dose and dose-rates are absorbed at 8, 15, 18 or 25~keV. This would help steer the future development of sources, as beamlines move to increasingly higher energy sources.\par 

This study showcases the importance of dose calculations for small molecule crystals, not only including the effect of varying photon energy, but a complete description of the experiment. Dose calculations can help inform the choice of beamline parameters for future experiments. Determining a dose limit allows for an understanding of the point at which the experimental dose should not be exceeded. However, dose alone is not the sole deciding factor governing which energy is chosen for an X-ray experiment, as shown in the results of the 25~keV dataset. Whilst a very minimal radiation dose is absorbed by the sample, the poor resolution diffraction patterns, meant that no accurate structural information could be extracted within the time frame of the experiment. This is attributed to the low photon flux and decrease in elastic scattering cross section which both contribute to the observed diffraction. In order to maximise the quality of data collected, the ideal energy would be that which corresponds to the maximum diffraction efficiency, ensuring the photon flux is high enough to obtain data of adequate signal quality within the experiment time, yet has a dose that falls just below the dose limit.\par

Developing robust routines to determine dose limits for small molecule crystals is also important. Established dose limits from protein crystallography, such as the Henderson and other experimental dose limits discussed in this work, cannot easily be translated to organometallic complexes, due to the absence of solvent-related damage processes in powder diffraction and other X-ray techniques. Here, we demonstrate that a diffraction dose limit, $D_{0.9}$, of 11.8~MGy is preferable with respect to previously proposed dose limits to describe structural sensitivities of the Rh complex studied, and other similar organometallic systems at room temperature. However, we suggest that the $D_{0.9}$ spectroscopic dose limit, proposed by Kubin~\textit{et al.}~and determined to be 7~MGy for this complex, should be given precedence in order to account for changes to local chemical environments, to which diffraction experiments are blind. As such, we note the limitations of defining universal dose limits in organometallic materials, due to the integral importance of dose rate on their stability and the dependence on whether specific or global damage is probed in the X-ray technique used.\par

To summarise, this work provides a detailed exploration of the influence of incident energy and related beam parameters on the crystal structure and attainable data quality in PXRD experiments. It provides a new approach to understanding and predicting dose-related behaviour in organometallic systems, which can help inform future experiments. Finally, it also clearly shows the importance of further investigations into the relationship of local and global damage in small molecule crystals as well as differing experimental environments across materials characterisation techniques.\par

\section*{Conflicts of interest}
There are no conflicts to declare.

\section*{Acknowledgements}
NKF acknowledges support from the Engineering and Physical Sciences Research Council (EP/L015277/1). JLD is funded on a Herchel Smith studentship. A.R. acknowledges the support from the
Analytical Chemistry Trust Fund for her CAMS-UK Fellowship. We acknowledge Diamond Light Source for time on beamline I11 under proposal EE19420. The authors also thank Dr.~Sarah Day, Beamline Scientist at beamline I11 and Dr.~Jonathan Spiers, Senior Detector Technician at Diamond Light Source for support in acquiring information on the photodiode used in flux calculations. We also thank Ed Rial of Helmholtz-Zentrum Berlin, for his useful insight into undulator mechanisms.



\balance


\bibliography{ref} 

\providecommand*{\mcitethebibliography}{\thebibliography}
\csname @ifundefined\endcsname{endmcitethebibliography}
{\let\endmcitethebibliography\endthebibliography}{}
\begin{mcitethebibliography}{49}
\providecommand*{\natexlab}[1]{#1}
\providecommand*{\mciteSetBstSublistMode}[1]{}
\providecommand*{\mciteSetBstMaxWidthForm}[2]{}
\providecommand*{\mciteBstWouldAddEndPuncttrue}
  {\def\EndOfBibitem{\unskip.}}
\providecommand*{\mciteBstWouldAddEndPunctfalse}
  {\let\EndOfBibitem\relax}
\providecommand*{\mciteSetBstMidEndSepPunct}[3]{}
\providecommand*{\mciteSetBstSublistLabelBeginEnd}[3]{}
\providecommand*{\EndOfBibitem}{}
\mciteSetBstSublistMode{f}
\mciteSetBstMaxWidthForm{subitem}
{(\emph{\alph{mcitesubitemcount}})}
\mciteSetBstSublistLabelBeginEnd{\mcitemaxwidthsubitemform\space}
{\relax}{\relax}

\bibitem[Holton(2009)]{Holton2009}
J.~M. Holton, \emph{Journal of Synchrotron Radiation}, 2009, \textbf{16},
  133--142\relax
\mciteBstWouldAddEndPuncttrue
\mciteSetBstMidEndSepPunct{\mcitedefaultmidpunct}
{\mcitedefaultendpunct}{\mcitedefaultseppunct}\relax
\EndOfBibitem
\bibitem[Garman(2010)]{Garman2010}
E.~F. Garman, \emph{Acta Crystallographica Section D Biological
  Crystallography}, 2010, \textbf{66}, 339--351\relax
\mciteBstWouldAddEndPuncttrue
\mciteSetBstMidEndSepPunct{\mcitedefaultmidpunct}
{\mcitedefaultendpunct}{\mcitedefaultseppunct}\relax
\EndOfBibitem
\bibitem[Taberman(2018)]{Taberman2018}
H.~Taberman, \emph{Crystals}, 2018, \textbf{8}, 157\relax
\mciteBstWouldAddEndPuncttrue
\mciteSetBstMidEndSepPunct{\mcitedefaultmidpunct}
{\mcitedefaultendpunct}{\mcitedefaultseppunct}\relax
\EndOfBibitem
\bibitem[Garman and Weik(2017)]{Garman2017}
E.~F. Garman and M.~Weik, \emph{Protein Crystallography. Methods in Molecular
  Biology}, Humana Press, New York, 2017, ch.~20, pp. 467--489\relax
\mciteBstWouldAddEndPuncttrue
\mciteSetBstMidEndSepPunct{\mcitedefaultmidpunct}
{\mcitedefaultendpunct}{\mcitedefaultseppunct}\relax
\EndOfBibitem
\bibitem[Morgan \emph{et~al.}(2018)Morgan, Kim, Blandy, Murray, Christensen,
  and Thompson]{Morgan2018}
L.~C. Morgan, Y.~Kim, J.~N. Blandy, C.~A. Murray, K.~E. Christensen and A.~L.
  Thompson, \emph{Chemical Communications}, 2018, \textbf{54}, 9849--9852\relax
\mciteBstWouldAddEndPuncttrue
\mciteSetBstMidEndSepPunct{\mcitedefaultmidpunct}
{\mcitedefaultendpunct}{\mcitedefaultseppunct}\relax
\EndOfBibitem
\bibitem[Christensen \emph{et~al.}(2019)Christensen, Horton, Bury, Dickerson,
  Taberman, Garman, and Coles]{Christensen2019}
J.~Christensen, P.~N. Horton, C.~S. Bury, J.~L. Dickerson, H.~Taberman, E.~F.
  Garman and S.~J. Coles, \emph{IUCrJ}, 2019, \textbf{6}, 703--713\relax
\mciteBstWouldAddEndPuncttrue
\mciteSetBstMidEndSepPunct{\mcitedefaultmidpunct}
{\mcitedefaultendpunct}{\mcitedefaultseppunct}\relax
\EndOfBibitem
\bibitem[Kubin \emph{et~al.}(2018)Kubin, Kern, Guo, K{\"{a}}llman, Mitzner,
  Yachandra, Lundberg, Yano, and Wernet]{Kubin2018}
M.~Kubin, J.~Kern, M.~Guo, E.~K{\"{a}}llman, R.~Mitzner, V.~K. Yachandra,
  M.~Lundberg, J.~Yano and P.~Wernet, \emph{Physical Chemistry Chemical
  Physics}, 2018, \textbf{20}, 16817--16827\relax
\mciteBstWouldAddEndPuncttrue
\mciteSetBstMidEndSepPunct{\mcitedefaultmidpunct}
{\mcitedefaultendpunct}{\mcitedefaultseppunct}\relax
\EndOfBibitem
\bibitem[George \emph{et~al.}(2008)George, Fu, Guo, Drury, Friedrich,
  Rauchfuss, Volkers, Peters, Scott, Brown, Thomas, and Cramer]{George2008}
S.~J. George, J.~Fu, Y.~Guo, O.~B. Drury, S.~Friedrich, T.~Rauchfuss, P.~I.
  Volkers, J.~C. Peters, V.~Scott, S.~D. Brown, C.~M. Thomas and S.~P. Cramer,
  \emph{Inorganica Chimica Acta}, 2008, \textbf{361}, 1157--1165\relax
\mciteBstWouldAddEndPuncttrue
\mciteSetBstMidEndSepPunct{\mcitedefaultmidpunct}
{\mcitedefaultendpunct}{\mcitedefaultseppunct}\relax
\EndOfBibitem
\bibitem[Arndt(1984)]{Arndt1984}
U.~W. Arndt, \emph{Journal of Applied Crystallography}, 1984, \textbf{17},
  118--119\relax
\mciteBstWouldAddEndPuncttrue
\mciteSetBstMidEndSepPunct{\mcitedefaultmidpunct}
{\mcitedefaultendpunct}{\mcitedefaultseppunct}\relax
\EndOfBibitem
\bibitem[Nave and Hill(2005)]{Nave2005}
C.~Nave and M.~A. Hill, \emph{Journal of Synchrotron Radiation}, 2005,
  \textbf{12}, 299--303\relax
\mciteBstWouldAddEndPuncttrue
\mciteSetBstMidEndSepPunct{\mcitedefaultmidpunct}
{\mcitedefaultendpunct}{\mcitedefaultseppunct}\relax
\EndOfBibitem
\bibitem[Polikarpov \emph{et~al.}(1997)Polikarpov, Teplyakov, and
  Oliva]{Polikarpov1997}
I.~Polikarpov, A.~Teplyakov and G.~Oliva, \emph{Acta Crystallographica Section
  D Biological Crystallography}, 1997, \textbf{53}, 734--737\relax
\mciteBstWouldAddEndPuncttrue
\mciteSetBstMidEndSepPunct{\mcitedefaultmidpunct}
{\mcitedefaultendpunct}{\mcitedefaultseppunct}\relax
\EndOfBibitem
\bibitem[M{\"{u}}ller \emph{et~al.}(2002)M{\"{u}}ller, Weckert, Zellner, and
  Drakopoulos]{Muller2002}
R.~M{\"{u}}ller, E.~Weckert, J.~Zellner and M.~Drakopoulos, \emph{Journal of
  Synchrotron Radiation}, 2002, \textbf{9}, 368--374\relax
\mciteBstWouldAddEndPuncttrue
\mciteSetBstMidEndSepPunct{\mcitedefaultmidpunct}
{\mcitedefaultendpunct}{\mcitedefaultseppunct}\relax
\EndOfBibitem
\bibitem[Weiss \emph{et~al.}(2005)Weiss, Panjikar, Mueller-Dieckmann, and
  Tucker]{Weiss2005}
M.~S. Weiss, S.~Panjikar, C.~Mueller-Dieckmann and P.~A. Tucker, \emph{Journal
  of Synchrotron Radiation}, 2005, \textbf{12}, 304--309\relax
\mciteBstWouldAddEndPuncttrue
\mciteSetBstMidEndSepPunct{\mcitedefaultmidpunct}
{\mcitedefaultendpunct}{\mcitedefaultseppunct}\relax
\EndOfBibitem
\bibitem[Shimizu \emph{et~al.}(2007)Shimizu, Hirata, Hasegawa, Ueno, and
  Yamamoto]{Shimizu2007}
N.~Shimizu, K.~Hirata, K.~Hasegawa, G.~Ueno and M.~Yamamoto, \emph{Journal of
  Synchrotron Radiation}, 2007, \textbf{14}, 4--10\relax
\mciteBstWouldAddEndPuncttrue
\mciteSetBstMidEndSepPunct{\mcitedefaultmidpunct}
{\mcitedefaultendpunct}{\mcitedefaultseppunct}\relax
\EndOfBibitem
\bibitem[Homer \emph{et~al.}(2011)Homer, Cooper, and Gonzalez]{Homer2011}
C.~Homer, L.~Cooper and A.~Gonzalez, \emph{Journal of Synchrotron Radiation},
  2011, \textbf{18}, 338--345\relax
\mciteBstWouldAddEndPuncttrue
\mciteSetBstMidEndSepPunct{\mcitedefaultmidpunct}
{\mcitedefaultendpunct}{\mcitedefaultseppunct}\relax
\EndOfBibitem
\bibitem[Donath \emph{et~al.}(2013)Donath, Brandstetter, Cibik, Commichau,
  Hofer, Krumrey, L{\"{u}}thi, Marggraf, M{\"{u}}ller, Schneebeli,
  Schulze-Briese, and Wernecke]{Donath2013}
T.~Donath, S.~Brandstetter, L.~Cibik, S.~Commichau, P.~Hofer, M.~Krumrey,
  B.~L{\"{u}}thi, S.~Marggraf, P.~M{\"{u}}ller, M.~Schneebeli,
  C.~Schulze-Briese and J.~Wernecke, \emph{Journal of Physics: Conference
  Series}, 2013, \textbf{425}, 062001\relax
\mciteBstWouldAddEndPuncttrue
\mciteSetBstMidEndSepPunct{\mcitedefaultmidpunct}
{\mcitedefaultendpunct}{\mcitedefaultseppunct}\relax
\EndOfBibitem
\bibitem[Dickerson and Garman(2019)]{Dickerson2019}
J.~L. Dickerson and E.~F. Garman, \emph{Journal of Synchrotron Radiation},
  2019, \textbf{26}, 922--930\relax
\mciteBstWouldAddEndPuncttrue
\mciteSetBstMidEndSepPunct{\mcitedefaultmidpunct}
{\mcitedefaultendpunct}{\mcitedefaultseppunct}\relax
\EndOfBibitem
\bibitem[Storm \emph{et~al.}(2021)Storm, Axford, and Owen]{Storm2021}
S.~L.~S. Storm, D.~Axford and R.~L. Owen, \emph{IUCrJ}, 2021, \textbf{8},
  2052--2525\relax
\mciteBstWouldAddEndPuncttrue
\mciteSetBstMidEndSepPunct{\mcitedefaultmidpunct}
{\mcitedefaultendpunct}{\mcitedefaultseppunct}\relax
\EndOfBibitem
\bibitem[Henderson(1990)]{Henderson1990}
R.~Henderson, \emph{Proceedings of the Royal Society of London. Series B:
  Biological Sciences}, 1990, \textbf{241}, 6--8\relax
\mciteBstWouldAddEndPuncttrue
\mciteSetBstMidEndSepPunct{\mcitedefaultmidpunct}
{\mcitedefaultendpunct}{\mcitedefaultseppunct}\relax
\EndOfBibitem
\bibitem[Owen \emph{et~al.}(2006)Owen, Rudino-Pinera, and Garman]{Owen2006}
R.~L. Owen, E.~Rudino-Pinera and E.~F. Garman, \emph{Proceedings of the
  National Academy of Sciences}, 2006, \textbf{103}, 4912--4917\relax
\mciteBstWouldAddEndPuncttrue
\mciteSetBstMidEndSepPunct{\mcitedefaultmidpunct}
{\mcitedefaultendpunct}{\mcitedefaultseppunct}\relax
\EndOfBibitem
\bibitem[Atakisi \emph{et~al.}(2019)Atakisi, Conger, Moreau, and
  Thorne]{Atakisi2019}
H.~Atakisi, L.~Conger, D.~W. Moreau and R.~E. Thorne, \emph{IUCrJ}, 2019,
  \textbf{6}, 1040--1053\relax
\mciteBstWouldAddEndPuncttrue
\mciteSetBstMidEndSepPunct{\mcitedefaultmidpunct}
{\mcitedefaultendpunct}{\mcitedefaultseppunct}\relax
\EndOfBibitem
\bibitem[Teng and Moffat(2000)]{Teng2000}
T.-y. Teng and K.~Moffat, \emph{Journal of Synchrotron Radiation}, 2000,
  \textbf{7}, 313--317\relax
\mciteBstWouldAddEndPuncttrue
\mciteSetBstMidEndSepPunct{\mcitedefaultmidpunct}
{\mcitedefaultendpunct}{\mcitedefaultseppunct}\relax
\EndOfBibitem
\bibitem[Nave and Garman(2005)]{Nave2005_2}
C.~Nave and E.~F. Garman, \emph{Journal of Synchrotron Radiation}, 2005,
  \textbf{12}, 257--260\relax
\mciteBstWouldAddEndPuncttrue
\mciteSetBstMidEndSepPunct{\mcitedefaultmidpunct}
{\mcitedefaultendpunct}{\mcitedefaultseppunct}\relax
\EndOfBibitem
\bibitem[Liebschner \emph{et~al.}(2015)Liebschner, Rosenbaum, Dauter, and
  Dauter]{Liebschner2015}
D.~Liebschner, G.~Rosenbaum, M.~Dauter and Z.~Dauter, \emph{Acta
  Crystallographica Section D Biological Crystallography}, 2015, \textbf{71},
  772--778\relax
\mciteBstWouldAddEndPuncttrue
\mciteSetBstMidEndSepPunct{\mcitedefaultmidpunct}
{\mcitedefaultendpunct}{\mcitedefaultseppunct}\relax
\EndOfBibitem
\bibitem[Ashfeld and Judd(2007)]{Ashfeld2007}
B.~L. Ashfeld and A.~S. Judd, \emph{Encyclopedia of Reagents for Organic
  Synthesis}, John Wiley {\&} Sons, Ltd, Chichester, 2007, ch. n/a, pp.
  1--13\relax
\mciteBstWouldAddEndPuncttrue
\mciteSetBstMidEndSepPunct{\mcitedefaultmidpunct}
{\mcitedefaultendpunct}{\mcitedefaultseppunct}\relax
\EndOfBibitem
\bibitem[Fernando \emph{et~al.}(2021)Fernando, Cairns, Murray, Thompson,
  Dickerson, Garman, Ahmed, Ratcliff, and Regoutz]{Fernando2021}
N.~K. Fernando, A.~B. Cairns, C.~A. Murray, A.~L. Thompson, J.~L. Dickerson,
  E.~F. Garman, N.~Ahmed, L.~E. Ratcliff and A.~Regoutz, \emph{The Journal of
  Physical Chemistry A}, 2021, \textbf{125}, 7473--7488\relax
\mciteBstWouldAddEndPuncttrue
\mciteSetBstMidEndSepPunct{\mcitedefaultmidpunct}
{\mcitedefaultendpunct}{\mcitedefaultseppunct}\relax
\EndOfBibitem
\bibitem[Cosier and Glazer(1986)]{Cosier1986}
J.~Cosier and A.~M. Glazer, \emph{Journal of Applied Crystallography}, 1986,
  \textbf{19}, 105--107\relax
\mciteBstWouldAddEndPuncttrue
\mciteSetBstMidEndSepPunct{\mcitedefaultmidpunct}
{\mcitedefaultendpunct}{\mcitedefaultseppunct}\relax
\EndOfBibitem
\bibitem[Ibers and Snyder(1962)]{Ibers1962}
J.~A. Ibers and R.~G. Snyder, \emph{Acta Crystallographica}, 1962, \textbf{15},
  923--930\relax
\mciteBstWouldAddEndPuncttrue
\mciteSetBstMidEndSepPunct{\mcitedefaultmidpunct}
{\mcitedefaultendpunct}{\mcitedefaultseppunct}\relax
\EndOfBibitem
\bibitem[{De Ridder} and Imhoff(1994)]{DeRidder1994}
D.~J.~A. {De Ridder} and P.~Imhoff, \emph{Acta Crystallographica Section C
  Crystal Structure Communications}, 1994, \textbf{50}, 1569--1572\relax
\mciteBstWouldAddEndPuncttrue
\mciteSetBstMidEndSepPunct{\mcitedefaultmidpunct}
{\mcitedefaultendpunct}{\mcitedefaultseppunct}\relax
\EndOfBibitem
\bibitem[Thompson \emph{et~al.}(2011)Thompson, Parker, Marchal, Potter, Birt,
  Yuan, Fearn, Lennie, Street, and Tang]{Thompson2011}
S.~P. Thompson, J.~E. Parker, J.~Marchal, J.~Potter, A.~Birt, F.~Yuan, R.~D.
  Fearn, A.~R. Lennie, S.~R. Street and C.~C. Tang, \emph{J. Synchrotron
  Radiat.}, 2011, \textbf{18}, 637--648\relax
\mciteBstWouldAddEndPuncttrue
\mciteSetBstMidEndSepPunct{\mcitedefaultmidpunct}
{\mcitedefaultendpunct}{\mcitedefaultseppunct}\relax
\EndOfBibitem
\bibitem[{Le Bail} \emph{et~al.}(1988){Le Bail}, Duroy, and
  Fourquet]{LeBail1988}
A.~{Le Bail}, H.~Duroy and J.~Fourquet, \emph{Materials Research Bulletin},
  1988, \textbf{23}, 447--452\relax
\mciteBstWouldAddEndPuncttrue
\mciteSetBstMidEndSepPunct{\mcitedefaultmidpunct}
{\mcitedefaultendpunct}{\mcitedefaultseppunct}\relax
\EndOfBibitem
\bibitem[Rietveld(1967)]{Rietveld1967}
H.~M. Rietveld, \emph{Acta Crystallographica}, 1967, \textbf{22},
  151--152\relax
\mciteBstWouldAddEndPuncttrue
\mciteSetBstMidEndSepPunct{\mcitedefaultmidpunct}
{\mcitedefaultendpunct}{\mcitedefaultseppunct}\relax
\EndOfBibitem
\bibitem[Rietveld(1969)]{Rietveld1969}
H.~M. Rietveld, \emph{Journal of Applied Crystallography}, 1969, \textbf{2},
  65--71\relax
\mciteBstWouldAddEndPuncttrue
\mciteSetBstMidEndSepPunct{\mcitedefaultmidpunct}
{\mcitedefaultendpunct}{\mcitedefaultseppunct}\relax
\EndOfBibitem
\bibitem[Coelho()]{TOPAS7}
A.~A. Coelho, \emph{{\textit{TOPAS Academic}: General Profile and Structure
  Analysis Software for Powder Diffraction Data, version 7.1 (Computer
  Software)}, Coelho Software: Brisbane, QLD, 2020}\relax
\mciteBstWouldAddEndPuncttrue
\mciteSetBstMidEndSepPunct{\mcitedefaultmidpunct}
{\mcitedefaultendpunct}{\mcitedefaultseppunct}\relax
\EndOfBibitem
\bibitem[Coelho(2018)]{Coelho2018}
A.~A. Coelho, \emph{J. Appl. Crystallogr.}, 2018, \textbf{51}, 210--218\relax
\mciteBstWouldAddEndPuncttrue
\mciteSetBstMidEndSepPunct{\mcitedefaultmidpunct}
{\mcitedefaultendpunct}{\mcitedefaultseppunct}\relax
\EndOfBibitem
\bibitem[Scheringer(1963)]{Scheringer1963}
C.~Scheringer, \emph{Acta Crystallographica}, 1963, \textbf{16}, 546--550\relax
\mciteBstWouldAddEndPuncttrue
\mciteSetBstMidEndSepPunct{\mcitedefaultmidpunct}
{\mcitedefaultendpunct}{\mcitedefaultseppunct}\relax
\EndOfBibitem
\bibitem[Zeldin \emph{et~al.}(2013)Zeldin, Gerstel, and Garman]{Zeldin2013}
O.~B. Zeldin, M.~Gerstel and E.~F. Garman, \emph{J. Appl. Crystallogr.}, 2013,
  \textbf{46}, 1225--1230\relax
\mciteBstWouldAddEndPuncttrue
\mciteSetBstMidEndSepPunct{\mcitedefaultmidpunct}
{\mcitedefaultendpunct}{\mcitedefaultseppunct}\relax
\EndOfBibitem
\bibitem[Bury \emph{et~al.}(2018)Bury, Brooks-Bartlett, Walsh, and
  Garman]{Bury2018}
C.~S. Bury, J.~C. Brooks-Bartlett, S.~P. Walsh and E.~F. Garman, \emph{Protein
  Sci.}, 2018, \textbf{27}, 217--228\relax
\mciteBstWouldAddEndPuncttrue
\mciteSetBstMidEndSepPunct{\mcitedefaultmidpunct}
{\mcitedefaultendpunct}{\mcitedefaultseppunct}\relax
\EndOfBibitem
\bibitem[Brooks-Bartlett \emph{et~al.}(2017)Brooks-Bartlett, Batters, Bury,
  Lowe, Ginn, Round, and Garman]{Brooks2017}
J.~C. Brooks-Bartlett, R.~A. Batters, C.~S. Bury, E.~D. Lowe, H.~M. Ginn,
  A.~Round and E.~F. Garman, \emph{Journal of Synchrotron Radiation}, 2017,
  \textbf{24}, 63--72\relax
\mciteBstWouldAddEndPuncttrue
\mciteSetBstMidEndSepPunct{\mcitedefaultmidpunct}
{\mcitedefaultendpunct}{\mcitedefaultseppunct}\relax
\EndOfBibitem
\bibitem[Owen \emph{et~al.}(2009)Owen, Holton, Schulze-Briese, and
  Garman]{Owen2009}
R.~L. Owen, J.~M. Holton, C.~Schulze-Briese and E.~F. Garman, \emph{Journal of
  Synchrotron Radiation}, 2009, \textbf{16}, 143--151\relax
\mciteBstWouldAddEndPuncttrue
\mciteSetBstMidEndSepPunct{\mcitedefaultmidpunct}
{\mcitedefaultendpunct}{\mcitedefaultseppunct}\relax
\EndOfBibitem
\bibitem[Thompson \emph{et~al.}(2009)Thompson, Parker, Potter, Hill, Birt,
  Cobb, Yuan, and Tang]{Thompson2009}
S.~P. Thompson, J.~E. Parker, J.~Potter, T.~P. Hill, A.~Birt, T.~M. Cobb,
  F.~Yuan and C.~C. Tang, \emph{Review of Scientific Instruments}, 2009,
  \textbf{80}, 075107\relax
\mciteBstWouldAddEndPuncttrue
\mciteSetBstMidEndSepPunct{\mcitedefaultmidpunct}
{\mcitedefaultendpunct}{\mcitedefaultseppunct}\relax
\EndOfBibitem
\bibitem[Arg(2013)]{Argonne2013}
\emph{{Compute X-ray Absorption}}, 2013,
  \url{https://11bm.xray.aps.anl.gov/absorb/absorb.php}\relax
\mciteBstWouldAddEndPuncttrue
\mciteSetBstMidEndSepPunct{\mcitedefaultmidpunct}
{\mcitedefaultendpunct}{\mcitedefaultseppunct}\relax
\EndOfBibitem
\bibitem[Scofield(1973)]{Scofield_1973}
J.~H. Scofield, \emph{{Theoretical Photoionization Cross Sections from 1 to
  1500 keV}}, U.s. atomic energy commission, divison of technical information
  extension technical report, 1973\relax
\mciteBstWouldAddEndPuncttrue
\mciteSetBstMidEndSepPunct{\mcitedefaultmidpunct}
{\mcitedefaultendpunct}{\mcitedefaultseppunct}\relax
\EndOfBibitem
\bibitem[Kalha \emph{et~al.}(2020)Kalha, Fernando, and Regoutz]{Dig_Sco_2020}
C.~Kalha, N.~K. Fernando and A.~Regoutz, \emph{{Digitisation of Scofield
  Photoionisation Cross Section Tabulated Data}}, 2020\relax
\mciteBstWouldAddEndPuncttrue
\mciteSetBstMidEndSepPunct{\mcitedefaultmidpunct}
{\mcitedefaultendpunct}{\mcitedefaultseppunct}\relax
\EndOfBibitem
\bibitem[Southworth-Davies \emph{et~al.}(2007)Southworth-Davies, Medina,
  Carmichael, and Garman]{Davies2007}
R.~J. Southworth-Davies, M.~A. Medina, I.~Carmichael and E.~F. Garman,
  \emph{Structure}, 2007, \textbf{15}, 1531--1541\relax
\mciteBstWouldAddEndPuncttrue
\mciteSetBstMidEndSepPunct{\mcitedefaultmidpunct}
{\mcitedefaultendpunct}{\mcitedefaultseppunct}\relax
\EndOfBibitem
\bibitem[Owen \emph{et~al.}(2012)Owen, Axford, Nettleship, Owens, Robinson,
  Morgan, Dor{\'{e}}, Lebon, Tate, Fry, Ren, Stuart, and Evans]{Owen2012}
R.~L. Owen, D.~Axford, J.~E. Nettleship, R.~J. Owens, J.~I. Robinson, A.~W.
  Morgan, A.~S. Dor{\'{e}}, G.~Lebon, C.~G. Tate, E.~E. Fry, J.~Ren, D.~I.
  Stuart and G.~Evans, \emph{Acta Crystallographica Section D Biological
  Crystallography}, 2012, \textbf{68}, 810--818\relax
\mciteBstWouldAddEndPuncttrue
\mciteSetBstMidEndSepPunct{\mcitedefaultmidpunct}
{\mcitedefaultendpunct}{\mcitedefaultseppunct}\relax
\EndOfBibitem
\bibitem[Murray and Garman(2002)]{Murray2002}
J.~Murray and E.~Garman, \emph{Journal of Synchrotron Radiation}, 2002,
  \textbf{9}, 347--354\relax
\mciteBstWouldAddEndPuncttrue
\mciteSetBstMidEndSepPunct{\mcitedefaultmidpunct}
{\mcitedefaultendpunct}{\mcitedefaultseppunct}\relax
\EndOfBibitem
\bibitem[Ravelli \emph{et~al.}(2002)Ravelli, Theveneau, McSweeney, and
  Caffrey]{Ravelli2002}
R.~B.~G. Ravelli, P.~Theveneau, S.~McSweeney and M.~Caffrey, \emph{Journal of
  Synchrotron Radiation}, 2002, \textbf{9}, 355--360\relax
\mciteBstWouldAddEndPuncttrue
\mciteSetBstMidEndSepPunct{\mcitedefaultmidpunct}
{\mcitedefaultendpunct}{\mcitedefaultseppunct}\relax
\EndOfBibitem
\bibitem[de~la Mora \emph{et~al.}(2020)de~la Mora, Coquelle, Bury, Rosenthal,
  Holton, Carmichael, Garman, Burghammer, Colletier, and Weik]{Mora2020}
E.~de~la Mora, N.~Coquelle, C.~S. Bury, M.~Rosenthal, J.~M. Holton,
  I.~Carmichael, E.~F. Garman, M.~Burghammer, J.-P. Colletier and M.~Weik,
  \emph{Proceedings of the National Academy of Sciences}, 2020, \textbf{117},
  4142--4151\relax
\mciteBstWouldAddEndPuncttrue
\mciteSetBstMidEndSepPunct{\mcitedefaultmidpunct}
{\mcitedefaultendpunct}{\mcitedefaultseppunct}\relax
\EndOfBibitem
\end{mcitethebibliography}


\providecommand*{\mcitethebibliography}{\thebibliography}
\csname @ifundefined\endcsname{endmcitethebibliography}
{\let\endmcitethebibliography\endthebibliography}{}
\begin{mcitethebibliography}{1}
\providecommand*{\natexlab}[1]{#1}
\providecommand*{\mciteSetBstSublistMode}[1]{}
\providecommand*{\mciteSetBstMaxWidthForm}[2]{}
\providecommand*{\mciteBstWouldAddEndPuncttrue}
  {\def\EndOfBibitem{\unskip.}}
\providecommand*{\mciteBstWouldAddEndPunctfalse}
  {\let\EndOfBibitem\relax}
\providecommand*{\mciteSetBstMidEndSepPunct}[3]{}
\providecommand*{\mciteSetBstSublistLabelBeginEnd}[3]{}
\providecommand*{\EndOfBibitem}{}
\mciteSetBstSublistMode{f}
\mciteSetBstMaxWidthForm{subitem}
{(\emph{\alph{mcitesubitemcount}})}
\mciteSetBstSublistLabelBeginEnd{\mcitemaxwidthsubitemform\space}
{\relax}{\relax}

\bibitem[Dubach and Guskov(2020)]{Dubach2020}
V.~R. Dubach and A.~Guskov, \emph{Crystals}, 2020, \textbf{10}, 580\relax
\mciteBstWouldAddEndPuncttrue
\mciteSetBstMidEndSepPunct{\mcitedefaultmidpunct}
{\mcitedefaultendpunct}{\mcitedefaultseppunct}\relax
\EndOfBibitem
\end{mcitethebibliography}
\bibliographystyle{rsc} 

\end{document}


\title{\textbf{Supplementary Information \\ Variability in X-ray induced effects in \ce{[Rh(COD)Cl]2} with changing experimental parameters}} 
\date{\vspace{-8ex}}
\maketitle

\begin{spacing}{1.25}

\noindent\large{Nathalie K. Fernando,\textit{$^{a}$} Hanna Bostr{\"o}m,\textit{$^{b}$} Claire A. Murray,\textit{$^{c}$} Robin L. Owen,\textit{$^{c}$} Amber L. Thompson,\textit{$^{d}$} Joshua L. Dickerson,\textit{$^{e}$} Elspeth F. Garman,\textit{$^{f}$} Andrew B. Cairns,\textit{$^{g}$} and Anna Regoutz\textit{$^{a}$}} \\

\noindent\textit{$^{a}$~Department of Chemistry, University College London, 20 Gordon Street, London, WC1H~0AJ, UK.}\\
\textit{$^{b}$~Max Planck Institute for Solid State Research, Heisenbergstra{\ss}e 1, 70569 Stuttgart, Germany.}\\
\textit{$^{c}$~Diamond Light Source Ltd, Diamond House, Harwell Science and Innovation Campus, Didcot, Oxfordshire, OX11~0DE, UK.}\\
\textit{$^{d}$~Chemical Crystallography, Chemistry Research Laboratory, University of Oxford, South Parks Road, Oxford OX1~3QR, UK.}\\
\textit{$^{e}$~MRC Laboratory of Molecular Biology, Francis Crick Avenue, Cambridge Biomedical Campus, Cambridge CB2 0QH, UK.}\\
\textit{$^{f}$~Department of Biochemistry, Dorothy Crowfoot Hodgkin Building, University of Oxford, South Parks Road, Oxford, OX1~3QU, UK.}\\
\textit{$^{g}$~Department of Materials, Imperial College London, Royal School of Mines, Exhibition Road, SW7~2AZ, UK.}\\

\clearpage

\section{Estimation of Dose}

\begin{figure}[h]
\centering
\includegraphics[width=0.5\textwidth]{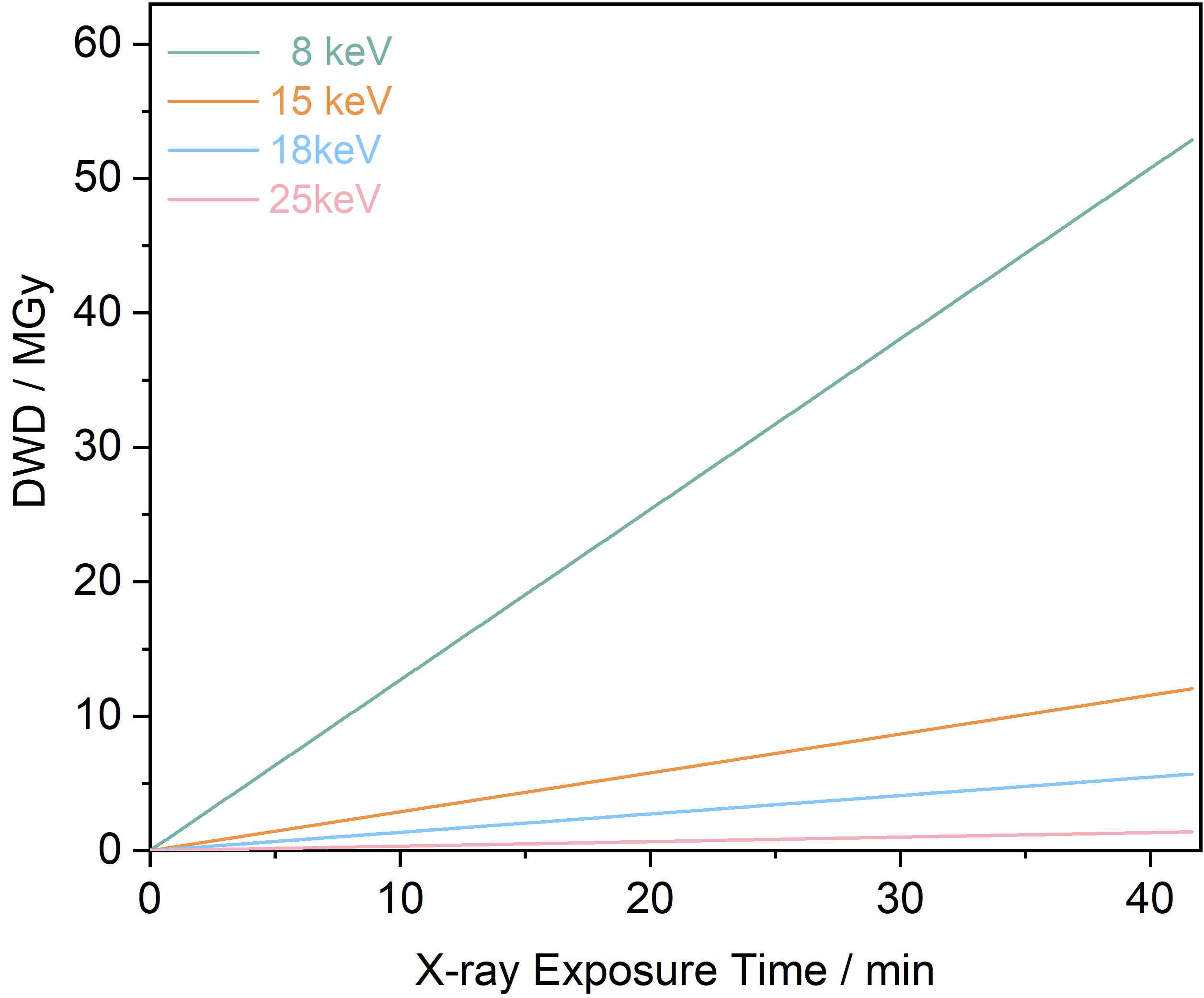}
\caption{The absorbed diffraction-weighted dose (DWD) calculated using RADDOSE-3D, as a function of X-ray exposure time for \ce{[Rh(COD)Cl]2}, at photon energies $h\nu$ of 8, 15, 18, and 25~keV.}
\label{fig:dose}
\end{figure}

\captionsetup[table]{labelfont=bf}

\begin{table*}[!ht]
    \caption{The parameters included in the RADDOSE-3D input file for \ce{[Rh(COD)Cl]2} used to estimate the doses for the varying photon energy setups in the powder XRD experiments at beamline I11 at Diamond Light Source, Didcot, UK.}\label{tab:raddose}
    \begin{tabular*}{\textwidth}{@{\extracolsep{\fill}}lllll}
        \hline
        RADDOSE parameter & 8~keV & 15~keV & 18~keV & 25~keV\\
        \hline
        Crystal type & cylinder & cylinder & cylinder & cylinder\\
        Crystal Dimension* /$\mu$m & 300 $\times$ 40000  & 300 $\times$ 40000 & 300 $\times$ 40000 & 300 $\times$ 40000\\
        PixelsPerMicron & 0.06 & 0.06 & 0.06 & 0.06\\
        Container material type & mixture & mixture & mixture & mixture \\
        Material mixture & pyrex & pyrex & pyrex & pyrex  \\
        Container thickness / $\mu$m & 10 & 10 & 10 & 10 \\
        Container density /   g\,cm$^{-3}$ & 2.23 & 2.23 & 2.23 & 2.23 \\
        Beam type & Gaussian & Gaussian & Gaussian & Gaussian \\
        Photon flux / ph\,s$^{-1}$ & 5.7 $\times$ 10$^{12}$ & 2.4 $\times$ 10$^{12}$ & 1.4 $\times$ 10$^{12}$ & 2.2 $\times$ 10$^{11}$\\
        FWHM / $\mu$m$^{2}$ & 2000 $\times$ 600  & 2000 $\times$ 600  & 2000 $\times$ 600  & 2000 $\times$ 600\\
        Energy / keV & 8 & 15 & 18 & 25\\
        Collimation type & rectangular & rectangular & rectangular & rectangular \\
        Collimation dimensions / $\mu$m$^{2}$ & 2500 $\times$ 800 & 2500 $\times$ 800 & 2500 $\times$ 800 & 2500 $\times$ 800\\
        \hline
        * diameter $\times$ length of capillary
    \end{tabular*}
\end{table*}

\clearpage

\section{PXRD Refinements}

\noindent The small, periodic discontinuities observed in the data, particularly evident in the 8~keV and 15~keV integrated intensity (peak area) plots as a function of time and dose (Figures~S3(a) and S3(d)) can be attributed to changes in beam current associated with the top-up mode of the synchrotron storage ring.

\begin{figure}[h]
\centering
\includegraphics[width=\textwidth]{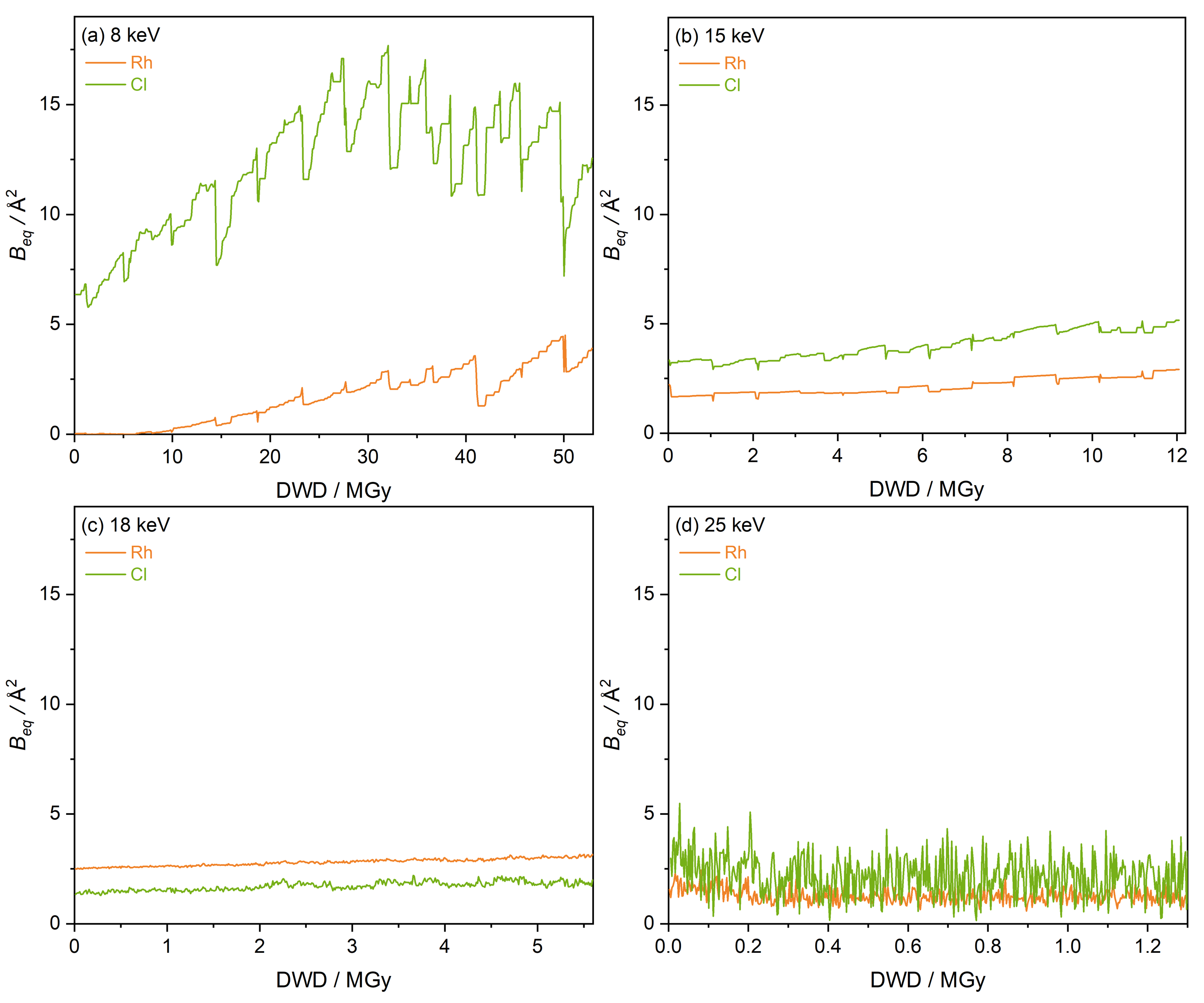}
\caption{The refined values of the isotropic atomic displacement parameter \textit{B$_{eq}$} for Rh and Cl at setups with a photon energy $h\nu$ of (a) 8~keV, (b) 15~keV, (c) 18~keV and (d) 25~keV as a function of diffraction-weighted dose (DWD). The carbon \textit{B$_{eq}$} values are fixed to a value of 2.0 throughout all 500 datasets.}
\label{fig:beq}
\end{figure}

\newpage

\begin{figure*}[h]
\centering
\includegraphics[width=\textwidth]{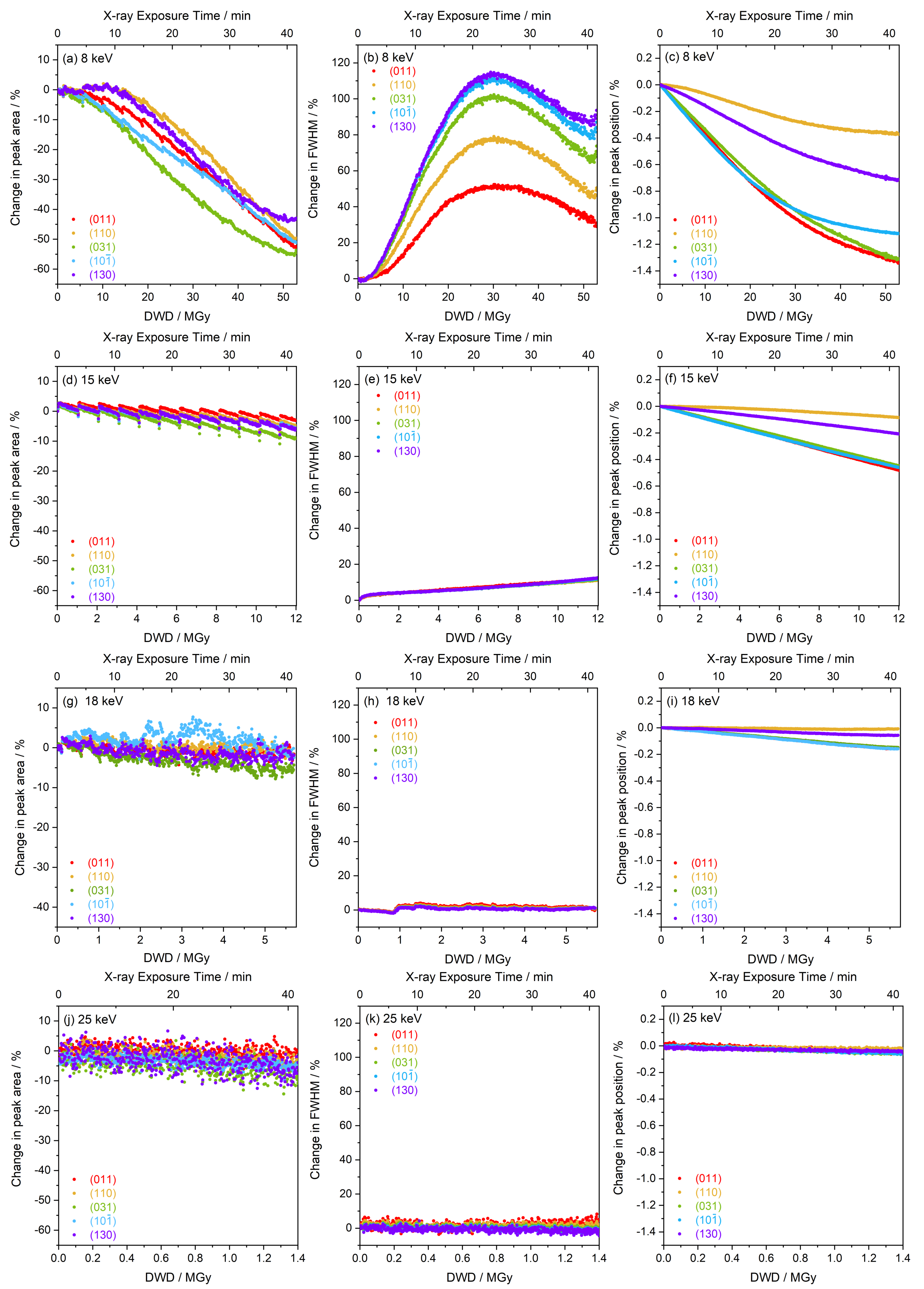}
\caption{The percentage change in peak intensity, FWHM, and peak shift for \ce{[Rh(COD)Cl]2} as a function of diffraction-weighted dose (DWD) and X-ray exposure time, obtained from Le Bail refinements, at setups with a photon energy $h\nu$ of (a) 8~keV, (b) 15~keV, (c) 18~keV and (d) 25~keV.}
\label{fig:lebail}
\end{figure*}

\clearpage

\section{Estimation of PXRD Resolution}

In single crystal macromolecular X-ray diffraction, discussion of resolution is routine, particularly in investigations of X-ray damage. However, this is not the case in powder X-ray diffraction studies. Although there is no direct equivalent, in order to bridge the gap between the two techniques, a somewhat crude approximation of PXRD resolution is made here. This was achieved by extracting the average error on the raw intensities, $\sigma$, obtained during Rietveld refinements. The diffraction peak at the furthest 2$\theta$ point, which has an intensity above the 2$\sigma$ threshold is considered to be the last resolvable peak,~\cite{Dubach2020} and converted to the $q$ `resolution' value. Tabulated below are the estimated $q$ values at each photon energy studied at the start ($t=5$\,s) and end ($t=41.7$\,min) of irradiation. It should be noted that due to the organic and monoclinic nature of the complex, individual peaks at high $2\theta$ values could not be resolved completely. Therefore the errors in the $d$-spacing are defined by the half-width of these broad peaks, which takes into account the $2\theta$ distribution of underlying peaks.

\begin{table*}[!ht]
\centering
  \caption{Estimation of the `resolution' of the diffraction experiments carried out in this work extracted from our X-ray diffraction data. The peak at the furthest 2$\theta$ point is shown for each start and end diffraction pattern at the four energies studied, which has an intensity above the 2$\sigma$ threshold, i.e. the last resolvable peak. This is tabulated with the associated Miller indices \textit{hkl}, d-spacing, \textit{d}, the error on \textit{d}, resolution, \textit{q} with its propagated error $\Delta$\textit{q} and the percentage change in \textit{q} from the minimum dose (start) to maximum dose (end) structure.}
  \label{tbl:dose_rate}
\begin{tabular}{llllllll}
\hline
$h\nu$ / keV & 2$\theta$ / $^{\circ}$ & \textit{h k l} & \textit{d} / \AA & $\Delta$\textit{d} & \textit{q} / \AA$^{-1}$  & $\Delta$\textit{q} & change in \textit{q} / \%\\
\hline
8\textsubscript{start}          & 60.1   & 1 15 2       & 1.55               & 0.01          & 4.06 & 0.03 & \multirow{2}{*}{$-61.8$}          \\
8\textsubscript{end}           & 22.1   & 0 3 2        & 4.04               & 0.04          & 1.56 & 0.02               \\ \hline
15\textsubscript{start}       & 48.1   & 7 6 0        & 1.014                 & 0.004            & 6.20 & 0.02 & \multirow{2}{*}{$-19.5$}            \\
15\textsubscript{end}           & 38.3   & 2 13 $\bar{5}$       & 1.26               & 0.01          & 4.98 & 0.05           \\ \hline
18\textsubscript{start}        & 35.5   & 6 7 $\bar{2}$        & 1.126               & 0.004          & 5.58 & 0.02 & \multirow{2}{*}{$-7.0$}            \\
18\textsubscript{end}          & 33.0   & 6 2 0       & 1.211               & 0.007          & 5.19 & 0.03         \\ \hline
25\textsubscript{start}        & 14.9   & 2 3 $\bar{4}$       & 1.91               & 0.04          & 3.29 & 0.08  & \multirow{2}{*}{$-4.5$}          \\
25\textsubscript{end}         & 14.2   & 1 11 $\bar{2}$       & 2.00               & 0.02          & 3.14 & 0.04 \\     
\hline
\end{tabular}
\end{table*}

\end{spacing}

\bibliography{ref} 
\bibliographystyle{rsc} 